\documentclass[aps,prb,twocolumn,superscriptaddress]{revtex4-2}
\pdfoutput=1
\usepackage{amsfonts}
\usepackage{amsmath}
\usepackage{graphicx}
\usepackage{dsfont}
\usepackage{diagbox}
\usepackage{array}
\usepackage{amssymb}
\usepackage{bbm}
\usepackage{float}
\usepackage{subfigure}
\usepackage{url}
\usepackage{hyperref}
\usepackage{array}
\usepackage{multirow}
\usepackage{bm}
\usepackage{booktabs}
\usepackage{bbm}
\usepackage{array}
\usepackage{color}
\usepackage[export]{adjustbox}
\usepackage{multirow}
\usepackage{makecell}
\usepackage{notes2bib}
\usepackage{float}

\newcommand{\PreserveBackslash}[1]{\let\temp=\\#1\let\\=\temp}
\newcolumntype{C}[1]{>{\PreserveBackslash\centering}p{#1}}
\newcolumntype{R}[1]{>{\PreserveBackslash\raggedleft}p{#1}}
\newcolumntype{L}[1]{>{\PreserveBackslash\raggedright}p{#1}}

\usepackage{color}

\newcommand{\ket}[1]{\lvert#1\rangle}

\begin{document}

\title{Characterization of topological phase transitions from a
non-Abelian topological state and its Galois conjugate
through condensation and confinement order parameters}
\author{Wen-Tao Xu}
\affiliation{University of Vienna, Faculty of Physics, Boltzmanngasse 5, 1090 Wien, Austria}
\author{Norbert Schuch}
\affiliation{University of Vienna, Faculty of Physics, Boltzmanngasse 5, 1090 Wien, Austria}
\affiliation{University of Vienna, Faculty of Mathematics, Oskar-Morgenstern-Platz 1, 1090 Wien, Austria}

\begin{abstract}
Topological phases exhibit unconventional order that cannot be detected
by any local order parameter. In the framework of Projected Entangled Pair
States (PEPS), topological order is characterized by an entanglement
symmetry of the local tensor that describes the model. This symmetry can
take the form of a tensor product of group representations (for quantum
double models $D(G)$ of a group $G$), or in the more general case a
correlated symmetry action in the form of a Matrix Product Operator (MPO),
that encompasses all string-net models, including those which are not
quantum double models. Among other things, these entanglement symmetries
allow for the succinct description of ground states and topological
excitations (anyons). Recently, the idea has been put forward to use those
symmetries and the anyonic objects they describe as order parameters for
probing topological phase transitions, and the applicability of this idea
has been demonstrated for Abelian groups. In this paper, we
extend this construction to the domain of non-Abelian models with MPO
symmetries,  and we use it to study the breakdown of topological order in the
double Fibonacci (DFib) string-net model and its Galois conjugate, namely the
non-hermitian double Yang-Lee (DYL) string-net model.  We start by showing how to
construct topological order parameters for condensation and deconfinement
of anyons using the MPO symmetries. Subsequently, we set up interpolations
from the DFib and the DYL model to the trivial phase, and we show that these
can be mapped to certain restricted solid on solid (RSOS) models,
which  are equivalent to the $((5\pm\sqrt{5})/2)$-state Potts model,
respectively. Moreover, the order parameter for condensation maps to the
RSOS order parameter.  The known exact solutions of the statistical models
subsequently allow us to locate
the critical points of the models, and to predict the critical exponents
for the order parameters from conformal filed theory. We complement this by numerical study
of the phase transitions, which fully confirms our theoretical
predictions; remarkably, we find that both models exhibit a duality
between the behavior of order parameters for condensation and
deconfinement.
\end{abstract}

\maketitle
\section{Introduction}
Landau's theory of spontaneous symmetry breaking is one of the
cornerstones of condensed matter physics. It captures the nature of phases
and phase transitions via local order parameters characterizing the
spontaneous breaking of the global symmetries of the system. This has been
challenged by the discovery of topologically ordered phases, which are non-trivial
phases without any global symmetries, and which therefore cannot be
characterized by local order parameters~\cite{Wen_2017_zoo}.
A prototypical example of a topological phase is realized by the toric
code model, and its generalizations based on
quantum doubles $D(G)$ of a finite group $G$
\cite{kitaev_toric_code_2003}.
 Although there is no global
symmetry, those models can be mapped to lattice gauge theories with
symmetry group $G$.  In addition to those phases, there exist a large
class of more exotic topological phases where anyons carry degrees of
freedom with irrational dimensions and whose gauge symmetry cannot be
described by group theory.   The concept of topological phases has even
been extended to non-hermitian systems~\cite{Galois_conjugate_2012}, which
exhibit new kinds of topological orders that cannot exist in hermitian
systems.

The fixed point wavefunctions of non-chiral topological phases can be
realized by so-called string-net models supporting anyonic excitations
\cite{Levin-Wen-string-net-2005}.
Both ground states of string-nets and excited states carrying anyonic
quasiparticles can be represented by projected entangled pair states
(PEPS)\cite{GuTensorNetwork_2009,buerschaper_explicit_2009},  which provide a description of the global wavefunction as a tensor
network built from local tensors. Here, the topological order is accompanied
by the presence of certain group
or Matrix Product Operator (MPO) symmetries in the entanglement degrees of
freedom of the tensor, which can be used to parametrize the ground space
manifold and anyonic excitations alike
\cite{peps_degeneracy_2010,sahinoglu:mpo-injectivity,MPO_algebra_2017}.
The description of topologically ordered systems as PEPS based on
entanglement symmetries suggests a natural way to construct and study
topological phase transitions within PEPS, by applying deformations to the
physical degrees of freedom that drive the system to a different phase
(such as a trivial product state)
\cite{Chen_Gu_Wen_LRE_2010,chen:topo-symmetry-conditions,schuch:topo-top,AnyonsCondensation_Z4_2018,xu_zhang_2018,zhu_gapless_2019,Xu_Zhang_Zhang_2020,zhang_xu_Wang_Zhang_2020}.
In this language, the entanglement symmetry in the tensor is preserved
throughout the path, but at some point, it no longer manifests itself in
topological order.

From the point of view of an effective theory, topological phase
transitions can be understood through the process of anyon condensation
and confinement: At a transition from a topological phase to one with a
lesser degree of topological order (such as a trivial phase), some anyons condense
into the ground state, and as a consequence, anyons that braid
non-trivially with condensed anyons must confine
\cite{Bais_Anyoncondensation_2009,burnell_anyon_2018}.
If the wavefunction of the system is given as a PEPS with entanglement
symmetries, such as for the interpolating families referenced above,
this process can be probed through topological order parameters which are
constructed at the entanglement level, and which probe the condensation
and deconfinement of anyons; notably, those order parameters can be used
to extract critical exponents which characterize the universal nature of
the phase transition, despite the lack of local order parameters~\cite{AnyonsCondensation_Z4_2017,AnyonsCondensation_Z4_2018,Iqbal_2020}.
However, up to now, the construction of condensation and deconfinement
order parameters, and the extraction of their universal scaling behavior
at criticality, has only been carried out for Abelian symmetry groups.

In this paper, we construct topological order parameters and extract the
critical exponents at the phase transitions for some of the most important
non-Abelian topological models, which furthermore cannot be constructed as
the quantum double of a group: The double Fibonacci (DFib) string-net
model~\cite{Levin-Wen-string-net-2005,Fibonacci_ladder_2009,J_Vidal_Golden_string_net_2013,schotte_tensor-network_2019,Vidal_2015},
and its Galois conjugate, the double Yang-Lee (DYL) string-net model,
which comes with a non-hermitian parent Hamiltonian
\cite{Galois_conjugate_2012, Micropic_Yang_Lee_2011,Galois_strange_correlator}.
For both models, we
set up a deformation that smoothly changes the model towards a trivial
product state, driving it through a topological phase transition. We show
how to construct order parameters for condensation and deconfinement,
using the MPOs underlying the entanglement symmetry of the tensors.
We continue by showing that the normalization of our deformed PEPS
wavefunctions can be mapped to the partition function of a restricted
solid on solid (RSOS) model associated with the Dynkin diagram $D_6$, where the order
parameter for condensation maps to the corresponding order parameter of
the RSOS model. That RSOS model, in turn, is known to map to
the $q$-state Potts model with $q=(2+\phi)$ and $q=(2-1/\phi)$ for the
DFib and the DYL model, respectively (where $\phi=(1+\sqrt5)/2$); for
the DFib model,
a duality to the same $q$ Potts model, albeit on the triangular lattice,
 has also been shown directly for
alternative
deformations~\cite{fidkowski_string_2009,fendley_topological_2008}.
This duality mapping allows us both to locate the exact critical point,
and to predict the critical exponents for the order parameter
(i.e., condensation); the self-duality of the model then suggests the same
critical exponents for the disorder parameters.

We supplement our analytical arguments with numerical study, which fully
matches the analytical findings, and in particular confirms the point that
the critical exponents for the disorder parameter (the anyon deconfinement
fraction) are the same as those for the order parameter (the anyon condensate
fraction), reinforcing the role played by the self-duality of the Potts
model.  Specifically, for the DFib model, we obtain that it is described
by the unitary minimal model with $c=14/15$, with critical exponents
$\eta=4/15$, $\nu=3/4$, and
$\beta=1/10$,
with $c$ the central charge and $\eta$, $\nu$,
and $\beta$ the exponents for correlations at criticality,
correlation length, and order parameter, respectively.
The DYL model is described by a non-unitary minimal model
with $c=8/35$, where the critical exponents are $\eta=8/35$, $\nu=7/6$,
and $\beta=2/15$.

The paper is organized as follows. Section II reviews the PEPS description
of the ground states and excited states of the string-net models. Section III
focuses on the analytic and numerical results of DFib string-net, and Sec.~IV
focuses on the analytic and numerical results of the DYL string-net.
Finally, we conclude in Sec.~V.

\section{PEPS representation for the string-net wavefunctions}
\subsection{String-net models with only one kind of strings}

A string-net model~\cite{Levin-Wen-string-net-2005} is specified by a set
of data $\{d_i, N_{ij}^{k},F_{tsu}^{ijk}\}$, where the indices can take
values in a set $\mathcal A$ of ``particles'', including the ``identity
particle'' $1\in \mathcal A$. Here, the \emph{fusion rule} $N_{ij}^k\in
\mathbb{N}$ counts the possible ways in which the particles $i$ and $j$ can
fuse to $k$, $d_i$ is the \emph{quantum dimension} of $i$, and the
$F_{tsu}^{ijk}$ must satisfy the so-called \emph{pentagon
equations}~\cite{Levin-Wen-string-net-2005}.
In this work, we are concerned with models with $\mathcal A=\{1,\tau\}$, which possess a non-trivial fusion rule
\begin{equation}
\label{eq:fusion}
  {1}\times{1}={1},\quad {1}\times {\tau}={\tau}\times
{1}={\tau},\quad {\tau}\times{\tau}={1}+{\tau}\ .
\end{equation}
This fusion rule allows for two solutions of the pentagon equations~\cite{Levin-Wen-string-net-2005}:
one unitary solution: the doubled Fibonacci (DFib) string-net model, and
one non-unitary solution: the doubled Yang-Lee (DYL) string-net model,
 which is the Galois conjugate of the DFib string-net model~\cite{Galois_conjugate_2012}.
For the DFib theory, $d_{1}=1$ and $d_{\tau}=\phi=(1+\sqrt{5})/2$, and
\begin{equation}\label{DFib_F_symbol}
  F^{\tau\tau i}_{\tau\tau j}=\frac{1}{\phi}\left(
                                              \begin{array}{cc}
                                                1 & \sqrt{\phi} \\
                                                \sqrt{\phi} & -1 \\
                                              \end{array}
                                            \right)_{ij}\ ,
\end{equation}
while all other entries allowed by the fusion rule $N_{ij}^k$ are
$1$, and $0$ otherwise.
The corresponding data for the DYL theory are obtained by replacing
$\phi$ with $\phi^\prime=-1/\phi=(1-\sqrt{5})/2$, resulting in
$d_{1}=1$, $d_{\tau}=\phi^{\prime}$ and the non-trivial entries of the $F$
symbol being
\begin{equation}\label{DYL_F_symbol}
  F^{\tau\tau i}_{\tau\tau j}=\frac{1}{\phi^{\prime}}\left(
                                              \begin{array}{cc}
                                                1 & \sqrt{\phi^{\prime}} \\
                                                \sqrt{\phi^{\prime}} & -1 \\
                                              \end{array}
                                            \right)_{ij}.
\end{equation}

The string-net model wavefunction on the honeycomb lattice is obtained
by assigning a degree of freedom $\{\ket0,\ket\tau\}$ to each vertex, and
constructing the wavefunction as a superposition of all configurations
that satisfy the fusion rule $N_{ij}^k$ across every vertex, with
amplitudes constructed from the quantum dimension $d_i$ and the $F$
symbol~\cite{Levin-Wen-string-net-2005}. These wavefunctions are exact ground
states of a local Hamiltonian which can also be constructed from the data
$\{d_i, N_{ij}^{k},F_{tsu}^{ijk}\}$~\cite{Levin-Wen-string-net-2005}.
An important difference between the Hamiltonians of DFib and DYL
string-net models is that the former is hermitian, while the latter is
non-hermitian; yet, the Hamiltonian of the DYL string-net model
has an entirely real energy
spectrum~\cite{Galois_conjugate_2012}.
The model is topologically ordered
and possesses anyonic excitations, which can be constructed from the
underlying particles $\mathcal A$ through doubling (by coupling a chiral and
an anti-chiral copy with particle content $\mathcal A$), and which we will
denote by $\{\bm 1, \bm \tau,\bm{\bar\tau},\bm b\}$, where $\bm\tau$ and
$\bm{\bar\tau}$ inherit the fusion rule \eqref{eq:fusion}, and $\bm
b=(\tau,\bar\tau)$ is the boson.

\subsection{PEPS representation for the ground states}

\begin{figure}
  \centering
  \includegraphics[width=8cm]{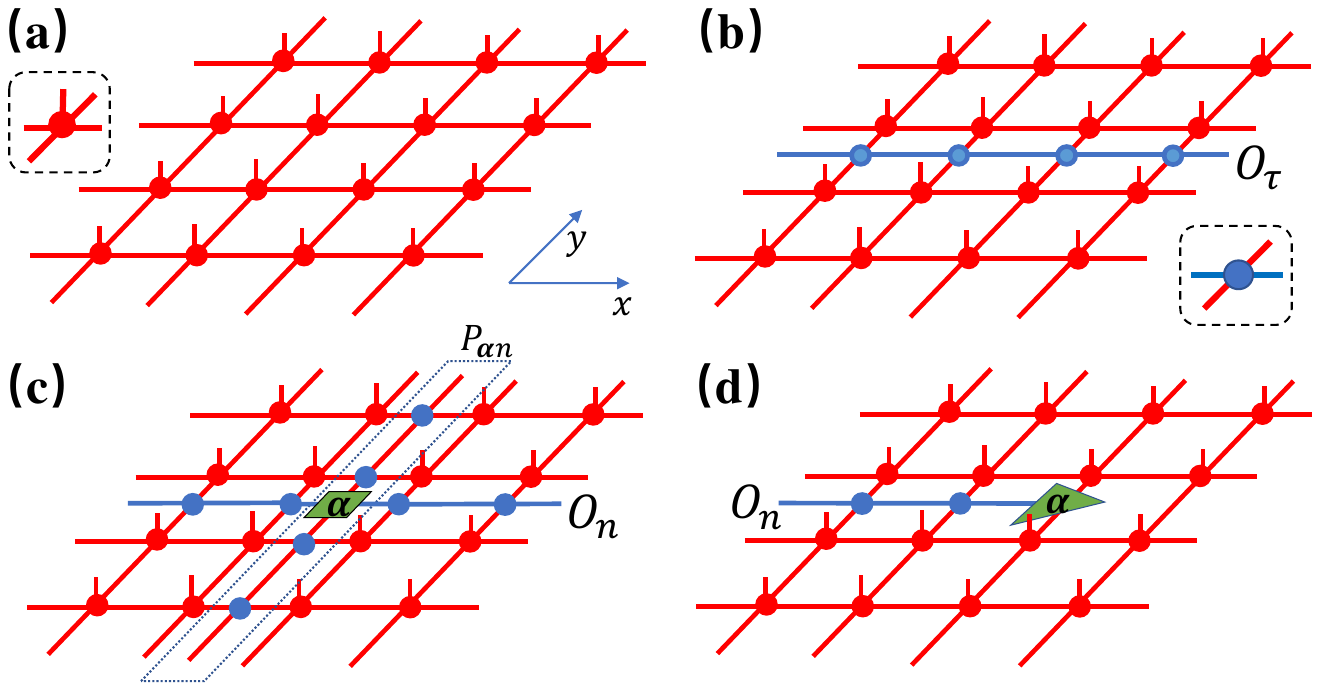}
  \caption{(a) A ground state of one of the topological models represented
by a PEPS; in the box is the local tensor generating the PEPS.  (b) A PEPS
with the non-trivial MPO $O_\tau$ inserted yields another ground state; in the
box is the local tensor generating the MPO.
   (c) The MES $|\bm{\alpha}\rangle$ with a well-defined horizontal
anyonic flux $\bm{\alpha}$ obtained by inserting the vertical idempotent
$P_{\bm{\alpha} n}$ into the PEPS in (a) or (b), where $n=1,\tau$ denotes
the type of
the horizontal MPO.
   (d) An excited state carrying an anyon $\bm{\alpha}$; the green
triangle is the rank-3 endpoint tensor
   $E_{\bm{\alpha}n}$.}\label{Fig_Tensor}
\end{figure}

The construction of Projected Entangled Pair States (PEPS) is illustrated
in Fig.~\ref{Fig_Tensor}(a): Here, each ball denotes a tensor, and the
legs denote indices. Connecting legs amounts to contracting (i.e.,
identifying and summing) the index.  Legs perpendicular to the $xy$  plane
are physical indices and legs parallel the $xy$ plane are the virtual
indices. Arranging local tensors in a two-dimensional grid as shown in the figure
(possibly on a different lattice)
and contracting their virtual indices give rise to the PEPS. The ground
state wavefunction of string-net models can be explicitly expressed as a
PEPS \cite{GuTensorNetwork_2009,buerschaper_explicit_2009}, whose local
tensors can be constructed from  $d_i$ and $F^{ijk}_{tsu}$. We describe
this construction in detail in
Appendix~\ref{Definitions_of_all_tensors}.
In the PEPS framework, the topological order is characterized by
virtual symmetries of the tensors described by
matrix product operators (MPOs)
\cite{peps_degeneracy_2010,MPO_algebra_2017,sahinoglu:mpo-injectivity}.  For
the DFib or DYL string-net model, there are two different MPOs describing
their order: One is the
trivial MPO $O_1\equiv\openone$, and the other is the non-trivial MPO $O_\tau$. Again, their
definitions are presented in Appendix~\ref{Definitions_of_all_tensors}.
The defining feature of these MPOs is that they can be freely moved. Thus,
inserting the non-trivial MPO $O_\tau$ into the
virtual level of the PEPS, as shown in Fig.~\ref{Fig_Tensor}(b),
results in another topologically degenerate ground state. Notice that
since $O_1$ equals the identity, a PEPS with
$O_1$ inserted is the same state as the one without inserting any MPO.

On a torus, the ground space of the DFib or DYL string-net model has a
four-fold topological degeneracy.  A canonical basis of the ground state
subspace is given by the \emph{minimally entangled states} (MESs)
$|\bm{\alpha}\rangle$, which have well defined anyonic flux $\bm{\alpha}$
along one direction of the torus \cite{MES}, where
$\bm{\alpha}=\bm{1},\bm{\tau},\bm{\bar{\tau}}$ and $\bm{b}$. As shown in
Fig.~\ref{Fig_Tensor}(c), in order to obtain an MES with a well-defined
horizontal anyonic flux, one needs to insert a vertical idempotent into the
PEPS~\cite{Anna_2020}. The idempotents come from the tube
algebra~\cite{Lan_Wen_2014,MPO_algebra_2017}, and consist of the MPO
tensors together with specific tensors inserted at the crossing point, see Appendix
\ref{Definitions_of_all_tensors} for their definitions.  There are four
central idempotents $P_{\bm{1}},P_{\bm{\tau}},P_{\bm{\bar{\tau}}}$ and
$P_{\bm{b}}$ of the tube algebra. By further specifying the type of the
horizontal MPO $O_n$ in Fig.~\ref{Fig_Tensor}(c) using a second
(non-boldface) subscript, it can be found that the first three central
idempotents are one dimensional:
$P_{\bm{1}}=P_{\bm{1}1},P_{\bm{\tau}}=P_{\bm{\tau}\tau},P_{\bm{\bar{\tau}}}=P_{\bm{\bar{\tau}}\tau}$,
but the last central idempotent is two dimensional:
$P_{\bm{b}}=P_{\bm{b}1}\oplus P_{\bm{b}\tau}$; see
Appendix~\ref{Definitions_of_all_tensors} for details.  Starting from the PEPS in
Fig.~\ref{Fig_Tensor}(a), we obtain the  MESs $|\bm{1}\rangle$ and
$|\bm{b}\rangle$ by inserting either the idempotent $P_{\bm{1}}$ or
$P_{\bm{b}1}$ in the vertical direction;  and starting from the PEPS in
Fig.~\ref{Fig_Tensor}(b), we obtain the MESs $|\bm{b}\rangle$,
$|\bm{\tau}\rangle$,
or $|\bm{\bar{\tau}}\rangle$ by inserting either the idempotent
$P_{\bm{b}\tau}$, $P_{\bm{\tau}}$, or $P_{\bm{\bar{\tau}}}$ in the
vertical direction.

\subsection{PEPS representation for the excited states}

The PEPS can also be used to represent the excited states of the
string-net models.  In order to describe the complete
anyonic excitations, the extended string-net models have been
proposed~\cite{extended_Levin_wen_2018}.  A plaquette term of an extended
string-net model is the trivial central idempotent of the tube algebra and
the anyonic excitations are projected out by other non-trivial central
idempotents acting around the plaquette. When applied in the
PEPS framework, an anyon excition is represented by a rank three
end point tensor $E_{\bm{\alpha}n}$ with an MPO
string attached to it~\cite{Schotte_2020_Fibonacci}, where the first
subscript
$\bm{\alpha}=\bm{1},\bm{b},\bm{\tau},\bm{\bar{\tau}}$ denotes the
anyon type, and the second subscript $n=1,\tau$ denotes the type of
attached MPO string, see Fig.~\ref{Fig_Tensor}(d). The end tensor
$E_{\bm{\alpha}n}\equiv E_{\bm{\alpha}n}(d_i,F_{tsu}^{ijk},R_{k}^{ij})$ is determined by
$d_i$ and $F_{tsu}^{ijk}$ together with a tensor $R_{k}^{ij}$, which
characterizes the
braiding statistics of $i$ and $j$ particles subjected to their fusion channel $k$,
see Appendix~\ref{Definitions_of_all_tensors}. According to the
definition of $E_{\bm{\alpha}n}$, there are five such end tensors:
$E_{\bm{1}1}$, $E_{\bm{\tau}\tau}$, $E_{\bm{\bar{\tau}}\tau}$,
$E_{\bm{b}1}$ and $E_{\bm{b}\tau}$, indicating that only trivial
(non-trivial) MPO strings can be attached to $\bm{1}$ ($\bm{\tau}$ and
$\bm{\bar{\tau}}$) excitations, while  both trivial and non-trivial MPO
strings can be attached to the $\bm{b}$ excitations.  Since the tensor
$E_{\bm{b}1}$ has the trivial MPO $O_1$ attached to it, this means that
the $\bm{b}$ excitation can be described by locally modifying the PEPS on
the virtual level, with no MPO string attached.

\section{DFib string-net}
\subsection{Deformed DFib string-net wavefunction}

\begin{figure}
  \centering
  \includegraphics[width=8cm]{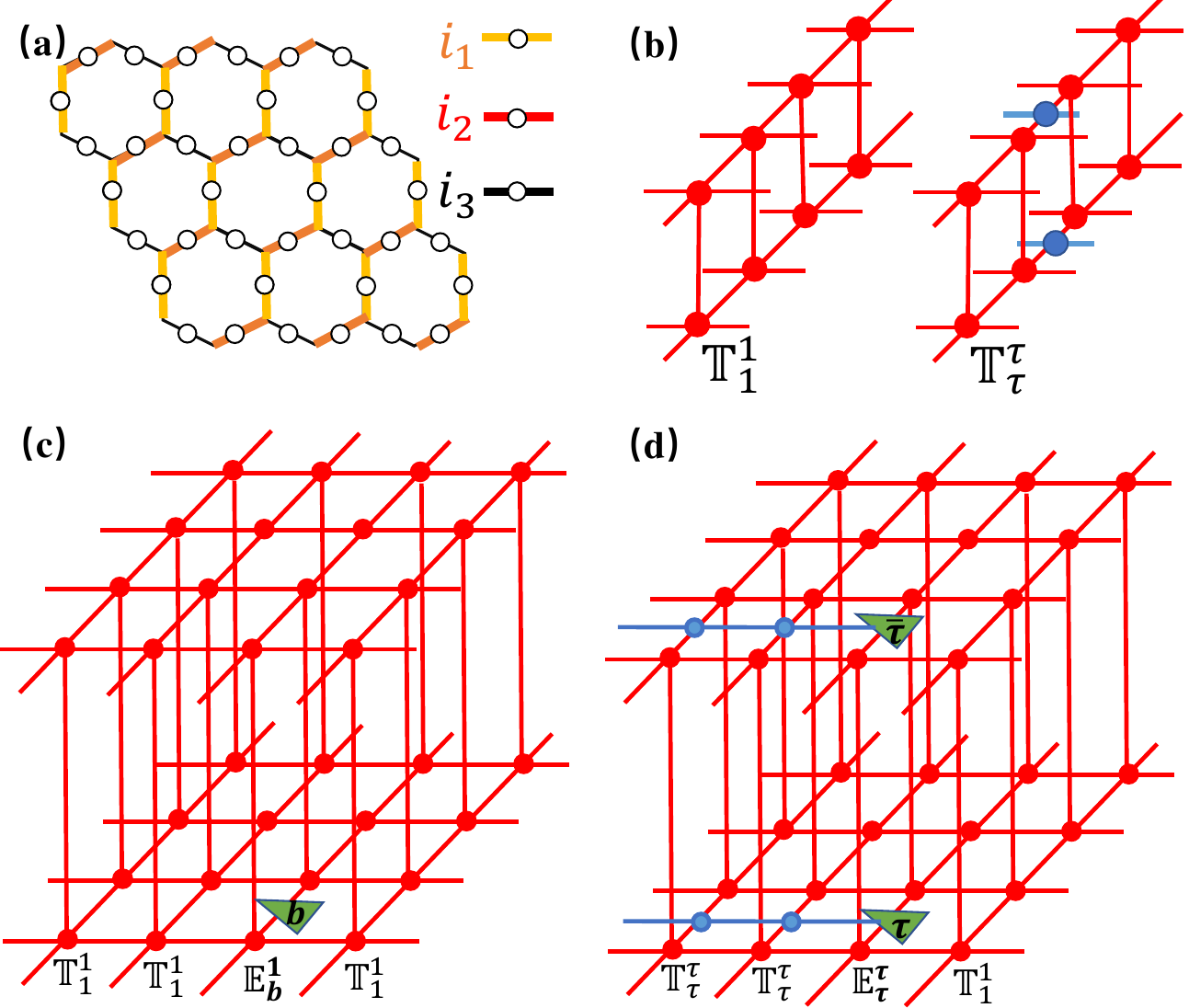}
  \caption{(a) The physical degrees of freedom of the model (open circles)
are located on the edges of a honeycomb lattice.
  The edges are classified into three sets $\{i_1\}$, $\{i_2\}$, and
$\{i_3\}$, according to their directions.
  (b) The transfer operators $\mathbb{T}_1^1$ and $\mathbb{T}_\tau^\tau$,
where the superscripts (subscripts) represent the
  types of MPO tensors inserted in bra (ket) layers.
  (c) The numerator of the condensate fraction
$\mathcal{C}_{\bm{b}}^{\bm{1}}$ represented as a tensor network.
  (d) The numerator of the deconfinement fraction
$\mathcal{C}_{\bm{\tau}}^{\bm{\tau}}$ represented as a tensor
network.}\label{figure_double_layer} \end{figure}

Let us now investigate what happens when we drive the DFib string-net model
into the trivial phase. To this end, we study a deformation of the DFib
string-net wavefunction, obtained by
imposing a string tension on the $\tau$ string (driving the system
towards  the topologically trivial vacuum state). Specifically, we add
different tensions $K_1$, $K_2$, and $K_3$ to the inequivalent edges of
the honeycomb lattice,
\begin{equation}
  |\Psi(K_1,K_2,K_3)\rangle=\prod_{i_1,i_2,i_3}e^\frac{K_1\sigma^z_{i_1}+K_2\sigma^z_{i_2}+K_3\sigma^z_{i_3}}{4}|\Psi_\text{DFib}\rangle,
\end{equation}
where $\sigma^z|1\rangle=|1\rangle$, $\sigma^z|\tau\rangle=-|\tau\rangle$, and
 $\{i_1\}, \{i_2\}$, $\{i_3\}$ denote the edges in each of the three
directions; see  Fig.~\ref{figure_double_layer}(a).
Importantly, since the deformation acts on the physical
degrees of freedom, the virtual MPO symmetry of the PEPS is preserved,
allowing us to construct the topological sectors and anyonic excitations
on top of the PEPS $|\Psi(K_1,K_2,K_3)\rangle$ as before.

In addition, the deformed wave function still has a frustration-free
parent Hamiltonian, which can be constructed by conjugating the
Hamiltonian of the DFib model (with the ground state energy of each term
shifted to zero) with the inverse of local
deformation~\cite{schuch:mps-phases,Condensation_driven_2017}.
Specifically, the Hamiltonian
$H=\sum_{r}h_r$ of the DFib string-net is a sum of the local positive
semi-definite projectors $h_r$ acting on the region $r$, and the deformation
matrix $\exp(K_t\sigma^z_{i_t}/4)$ with $t=1,2,3$ is also a local positive
definite operator, so that the parent Hamiltonian of the deformed wave function
is
\begin{equation}
\label{eq:ham-deformed-dfib}
H(K)=\sum_{r}P_r^{-1}h_rP_r^{-1}\ ,
\end{equation}
where $P_r=\prod_{i_t\in
r}\exp(K_t\sigma^z_{i_t}/4)$.
The possible quantum critical points of
this Rokhsar-Kivelson type Hamiltonian are the so-called conformal quantum
critical points \cite{ardonne_2004}, where all equal-time correlation
functions are described by two-dimensional conformal field theories
(CFTs), which can be extracted from the transfer operators of the
wavefunction norms at the critical points.

When $K_1=K_2=K_3=K$, it has been shown that the norm of the deformed
wavefunction can be mapped to the partition function of the isotropic
$(\phi+2)$-state Potts model on the dual triangular
lattice\cite{fidkowski_string_2009,fendley_topological_2008}. As the
string-tension $K$ increases, there is a phase transition from the
topological phase to the non-topological phase, where the position of the
critical point and the CFT describing it are known from the exact solution
of the Potts model.

We will in the following consider a different case, namely $K_1=K_2=K$,
$K_3=0$. As we prove in Appendix~\ref{Map_to_RSOS}
by also taking the virtual degrees of freedom in the PEPS into account,
the norm of the deformed PEPS equals the partition function of an
RSOS model on the square lattice associated with the $D_6$ Dynkin
diagram\cite{pasquier_Dynkin_RSOS}. Moreover, it has been shown that the
partition functions of RSOS models associated with $D$ type Dynkin
diagrams and the partition functions of Potts models are
equivalent\cite{he_2020_geometrical}.  Therefore, we can conclude that the
norm of the deformed wavefunction
maps to the partition function of the
$q=(\phi+2)$-state Potts model on the \emph{square} lattice:
\begin{equation}\label{quantum-classical-mapping}
  \langle\Psi(K,K,0)|\Psi(K,K,0)\rangle\propto\mathcal{Z}(q=\phi+2,K)\ .
\end{equation}

If we think of the virtual degrees of freedom of the PEPS as the
``Potts spins'' and the physical degrees of freedom as their ``domain
walls'' -- a picture which e.g.\ underlies the well-known mapping between
$\mathbb Z_N$ quantum doubles and Potts or clock models with $q=N$ -- we see that
the disordered phase of the Potts model corresponds to the topological
phase at small $K$ (where domain walls strongly fluctuate), while the
ordered phase of the Potts model corresponds to the trivial phase at large
$K$ (where domain wall fluctuations are supressed).

Under the self-duality of the the square lattice Potts models, the
partition function $\mathcal{Z}(q,K)$ of the $q$-state Potts model is
mapped to $\mathcal{Z}(q,K^\star)$, where $K$ and $K^\star$ satisfy
$(e^K-1)(e^{K^\star}-1)=q$~\cite{wu_potts_1982}.  The critical point of
the $q$-state Potts model (for $q\leq4$, as is the case here) is known to
be located at the self-dual point.  Its position $e^{K_c}$ and the central
charge $c$ of the CFT describing it are~\cite{saleur_1991}
\begin{equation}
\label{central_charge}
e^{K_c}=1+\sqrt{q}\ ,\quad c=1-6/[\delta(\delta-1)]\ ,
\end{equation}
where $q=4\cos^2(\pi/\delta)$.
When $q=\phi+2$, $e^{K_c}\approx2.9021$, and the CFT describing the
critical point is the unitary minimal model with the central charge
$c=14/15$.

\subsection{Condensate and deconfinement fractions and correlation functions}

Topological phase transitions in PEPS can be characterized by order
parameters constructed from the anyonic excitations of the theory, which
measure the condensation and deconfinement of the anyons,
respectively~\cite{AnyonsCondensation_Z4_2017,AnyonsCondensation_Z4_2018}.
Only anyons with bosonic self-statistics can
condense~\cite{Bais_Anyoncondensation_2009,burnell_anyon_2018}; for the DFib model, this is
only the $\bm b$ anyon. We define
its condensate fraction as
\begin{equation}
 \mathcal{F}_{\bm{b}}^{\bm{1}}=\frac {\langle
\bm{1}|\dot{\bm{b}}_i\rangle}{\langle \bm{1}|\bm{1}\rangle}\ ,
\end{equation}
where $|\bm{1}\rangle$ is the minimally entangled ground state with a
trivial anyon flux, and $|\dot{\bm{b}}_i\rangle$ is an excited state
obtained by creating a $\bm{b}$ anyon at position $i$ on top of
$|\bm{1}\rangle$. Since the $\bm{b}$ anyons created using $E_{\bm{b}1}$
and $E_{\bm{b}\tau}$ are equivalent, we choose to create the  $\bm{b}$ anyon using
$E_{\bm{b}1}$ for simplicity, since this does not require to attach an MPO
string on the virtual level of PEPS,
see Fig.~\ref{figure_double_layer}(c).
In the topological phase, $|\dot{\bm{b}}_i\rangle$ is a well defined excited state which is orthogonal to
the ground state $|\bm{1}\rangle$, and thus, we expect that the condensate fraction is zero.
In the topologically trivial phase, which is obtained by condensing the
$\bm{b}$ anyon, we correspondingly expect a non-zero condensate fraction.

In the topologically trivial phase, the condensation of $\bm{b}$ must be
accompanied by the confinement of the $\bm{\tau}$ and $\bm{\bar{\tau}}$
anyons, because they have non-trivial mutual statistics with
$\bm{b}$~\cite{Bais_Anyoncondensation_2009,burnell_anyon_2018}.
The deconfinement fraction of $\bm{\tau}$ can be defined as
\begin{equation}
   \mathcal{F}_{\bm{\tau}}^{\bm{\tau}}=\frac{\langle\dot{\bm{\tau}}_i|\dot{\bm{\tau}}_i\rangle}{\langle\bm{1}|\bm{1}\rangle}\ ,
\end{equation}
where $|\dot{\bm{\tau}}_i\rangle$ is obtained by inserting an end tensor $E_{\bm{\tau}}$ with
a semi-infinite non-trivial MPO string attached, see
Fig.~\ref{figure_double_layer}(d).  Because $\bm{\tau}$ anyons are
deconfined (confined) in the topological (non-topological) phase, we
expect the
deconfinement fraction to be nonzero (zero).

These anyonic order parameters for topological phases also open up a new
perspective on order and disorder parameters for the corresponding Potts
model with non-integer $q$:
As we discussed before, the topological and trivial phases of the
deformed DFib PEPS are mapped to the disordered and ordered phases of the
(non-integer) $q=(\phi+2)$-state Potts model.
Thus, we can interpret the condensate fraction $\mathcal F_{\bm{b}}^{\bm
1}$ as an order parameter for the Potts model, since it is non-zero (zero)
in the ordered (disordered) phase.  On the other hand, the deconfinement
fraction $\mathcal F_{\bm{\tau}}^{\bm\tau}$ can be interpreted as a
disorder parameter~\cite{fradkin2017disorder} for the $q=(\phi+2)$-state
Potts model, since it is non-zero (zero) in the disordered (ordered)
phase.  For the ordered phase, we strengthen this connection by proving in
Appendix \ref{Map_to_RSOS} that the condensate fractions
$\mathcal{F}_{\bm{b}}^{\bm{1}}$ and $\mathcal{F}_{\bm{1}}^{\bm{b}}$ are
equivalent to the expectation values of the local order parameters of the
RSOS model\cite{pasquier_1987_operator}. Together with the mapping
between RSOS and Potts models, this further strengthens the interpretation
of the condensate fraction as an order parameter for the Potts model.

In addition to the condensate and deconfinement fractions, the
correlation functions between pairs of anyons also contain useful
information; in fact, the order parameters above can (just as any order
parameter) be seen as the square root of the asymptotic value of an
underlying correlation function. Specifically,  the correlation function
underlying the condensate fraction of $\bm{b}$ anyons can be defined by
creating a pair of $\bm{b}$ anyons in the ket layer and computing their
overlap with the trivial MES,
${\langle\bm{1}|\dot{\bm{b}}_i\dot{\bm{b}}_j\rangle}/{\langle\bm{1}|\bm{1}\rangle}$.
Specifically, in the topological phase, it will
display an exponential decay
\begin{equation}\label{cond_corr}
  \mathcal{C}_{\bm{b}}^{\bm{1}}(|i-j|)=\frac{\langle\bm{1}|\dot{\bm{b}}_i\dot{\bm{b}}_j\rangle}{\langle\bm{1}|\bm{1}\rangle}\propto
\exp{\left(-\frac{|i-j|}{\xi^{\bm{1}}_{\bm{b}}}\right)}\ ,
\end{equation}
where the inverse correlation length $1/\xi^{\bm{1}}_{\bm{b}}$ can be interpreted as
the ``anyon mass gap'' \cite{Iqbal_2020}, while in the trivial phase, it will
converge to $\mathcal |F^{\bm 1}_{\bm b}|^2>0$.
Similarly, the correlation function underlying the confinement of $\bm{\tau}$ can be
constructed by creating a pair of $\bm{\tau}$ anyons and considering
their norm,
${\langle
\dot{\bm{\tau}}_i\dot{\bm{\tau}}_j|\dot{\bm{\tau}}_i\dot{\bm{\tau}}_j\rangle}/
{\langle \bm{1}|\bm{1}\rangle}$, which we expect to
decay exponentially in the topologically trivial phase,
\begin{equation}\label{conf_corr}
  \mathcal{C}_{\bm{\tau}}^{\bm{\tau}}(|i-j|)=\frac {\langle \dot{\bm{\tau}}_i\dot{\bm{\tau}}_j|\dot{\bm{\tau}}_i\dot{\bm{\tau}}_j\rangle}{\langle \bm{1}|\bm{1}\rangle}\propto \exp{\left(-\frac{|i-j|}{\xi_{\bm{\tau}}^{\bm{\tau}}}\right)},
\end{equation}
with $\xi_{\bm{\tau}}^{\bm{\tau}}$ the confinement length scale; again, in
the topological phase, this will converge to $|\mathcal
F_{\bm\tau}^{\bm\tau}|^2$.
Again, both of these correlation lengths can also be interpreted in the
Potts model as the correlation lengths corresponding to the order and
disorder parameter, respectively.

Finally, let us define a trivial correlation function
\begin{equation}\label{trivial_corr}
\mathcal{C}_{\bm{1}}^{\bm{1}}(|i-j|)=\frac{\langle\bm{1}|\sigma_i^z\sigma_j^z|\bm{1}\rangle}{\langle\bm{1}|\bm{1}\rangle}-\frac{\langle\bm{1}|\sigma_i^z|\bm{1}\rangle}{\langle\bm{1}|\bm{1}\rangle}\frac{\langle\bm{1}|\sigma_j^z|\bm{1}\rangle}{\langle\bm{1}|\bm{1}\rangle},
\end{equation}
where the operators $\sigma_i^z$ act on the physical level of the PEPS
instead of on the virtual level.
Since the internal energy of the Potts model is
$-\frac{1}{\mathcal{Z}}\frac{\partial
\mathcal{Z}} {\partial K}=-\sum_{i}\langle \sigma^z_i\rangle$,
$\sigma_i^z$ is the energy operator and the above correlation function can
be interpreted as a correlation function of energy operators of the
Potts model.

\subsection{Prediction of the critical expotents from CFT}

\begin{figure*}
  \centering
  \includegraphics[width=16cm]{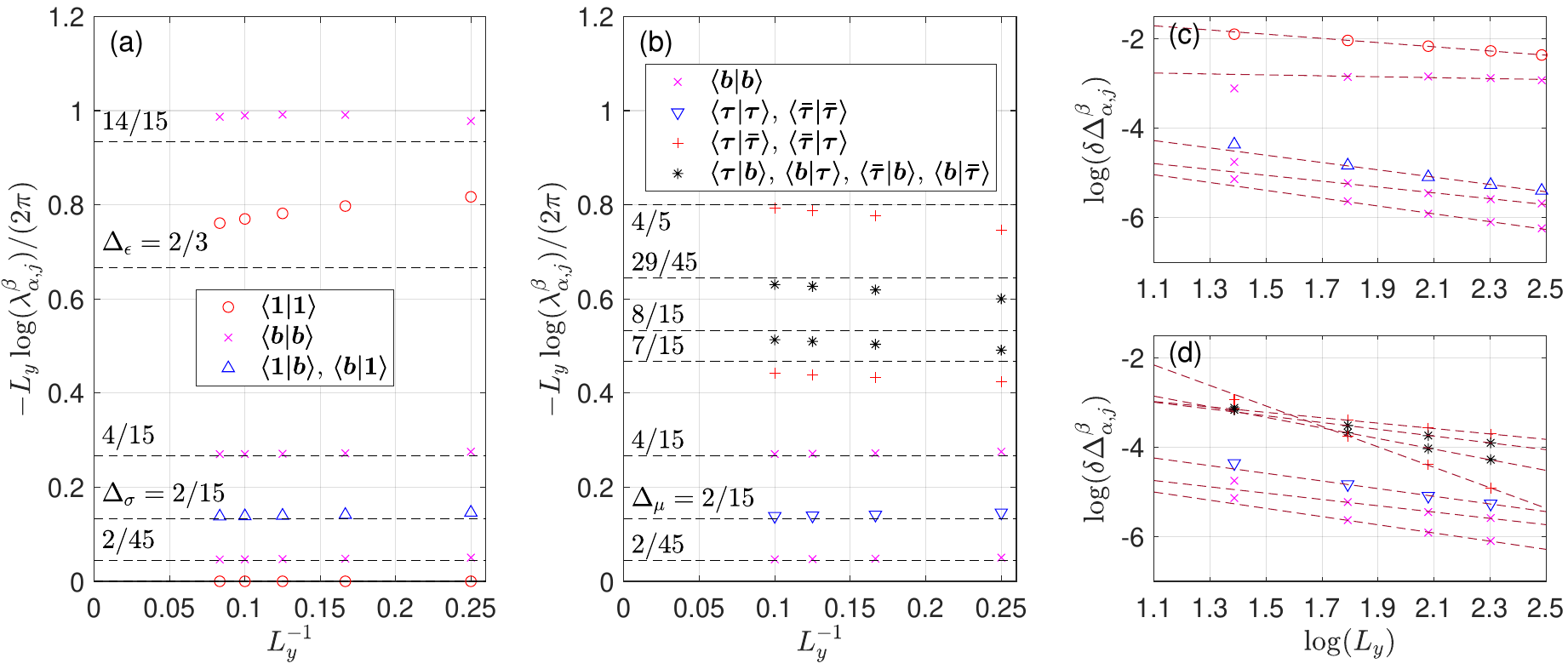}
  \caption{The numerically extracted scaling dimensions of primary fields
for the DFib case. (a)-(b) The scaling dimensions
$-L_y\log(\lambda_{\bm{\alpha},j}^{\bm{\beta}})/2\pi$ numerically
extracted from $\mathbb{T}_1^1$ (left) and $\mathbb{T}_{\tau}^{\tau}$
(right) on a cylinder with circumference $L_y$, where
$\lambda_{\bm{\alpha},j}^{\bm{\beta}}$ are rescaled eigenvalues,
classified into different topological sectors
$\langle\bm{\alpha}|\bm{\beta}\rangle$. The black dashed lines are the
predictions from CFT.
 $\Delta_\sigma$, $\Delta_\epsilon$ and $\Delta_\mu$
label the scaling dimensions for order parameter, energy, and disorder
operators, respectively. (c)-(d) The finite-size corrections
$\delta\Delta_{\bm{\alpha},j}^{\bm{\beta}}(L_y)=|-L_y\log(\lambda_{\bm{\alpha},j}^{\bm{\beta}})/2\pi-\Delta_{\bm{\alpha},j}^{\bm{\beta}}|$
of the scaling dimensions extracted from $\mathbb{T}_{1}^{1}$ (above) and
$\mathbb{T}_{\tau}^{\tau}$ (below) vanish algebraically with $L_y$, where
$\Delta_{\bm{\alpha},j}^{\bm{\beta}}$ are the exact values predicted from
CFT.}\label{DFib_CFT_spec} \end{figure*}
Since the condensate and deconfinement fractions and the corresponding
anyon correlation functions are interpreted as order and disorder
parameters and correlation functions of the $(\phi+2)$-state Potts model,
respectively, we expect a scaling
behavior
\begin{equation}\label{exponents_near_K_c}
   \mathcal{F}_{\bm{b}}^{\bm{1}}\propto t_{+}^{\beta}, \quad \mathcal{F}_{\bm{\tau}}^{\bm{\tau}}\propto t_{-}^{\beta^\star},\quad 1/\xi^{\bm{1}}_{\bm{b}}\propto t_{-}^{\nu}, \quad 1/\xi_{\bm{\tau}}^{\bm{\tau}}\propto t_{+}^{\nu^\star}
\end{equation}
near the critical point $K_c$,
where $t_{\pm}=\pm(e^{K}-e^{K_c})$, and $\beta$, $\beta^{\star}$, $\nu$
and $\nu^{\star}$ are critical exponents to be determined. In addition, at
the critical point the correlation functions will decay algebraically
with critical exponents $\eta$ and $\eta^\star$:
\begin{equation}\label{exponents_at_K_c}
\mathcal{C}_{\bm{b}}^{\bm{1}}(|i-j|)\propto|i-j|^{-\eta},\quad
 \mathcal{C}_{\bm{\tau}}^{\bm{\tau}}(|i-j|)\propto|i-j|^{-\eta^\star}.
\end{equation}

At first, let's consider the critical exponents $\eta$, $\nu$ and $\beta$
of the condensate fraction. From CFT, it is
well-know that at the critical point the critical exponents of the various
correlation functions are determined by the scaling dimensions of the
corresponding primary fields. Since the
condensate fractions are
mapped to the expectation values of the RSOS order parameters, whose
scaling dimension
$\Delta_\sigma=2/15$
is known \cite{pasquier_1987_operator}, we have $\eta=2\Delta_\sigma=4/15$.
The scaling dimension of the RSOS order parameters
also coincides with the magnetic exponent of the Potts
model \cite{Potts_scaling_dims}. Moreover, at the critical point, the
trivial correlation function decays algebraically:
$\mathcal{C}_{\bm{1}}^{\bm{1}}(|i-j|)\propto|i-j|^{-2\Delta_{\epsilon}}$,
where $\Delta_\epsilon$ is the scaling dimension of the Potts energy
operator. From Ref.~\cite{Potts_scaling_dims}, we know that
$\Delta_{\epsilon}=2/3$ for the $(\phi+2)$-state Potts model. To
double-check these findings, we have also numerically extracted the
scaling dimensions from the transfer operator spectrum of $\mathbb T_{\bm
1}^{\bm 1}$ on finite cylinders, labeled by
different topological sectors; see Appendix~\ref{corr_form_factor} for
details. The results, shown in Fig.~\ref{DFib_CFT_spec} (a), are in full
agreement with the analytical results.
Additionally using that from the scaling hypothesis~\cite{CFT}, we have
$\Delta_{\epsilon}=2-1/\nu, \beta=\eta\nu/2$,
and we can derive all three critical exponents:
 $\eta=4/15$, $\nu=3/4$, $\beta=1/10$.

Next, we consider the critical exponents $\eta^\star$, $\nu^\star$, $\beta^\star$.
The critical exponent $\eta^\star$ should be determined by the scaling dimension $\Delta_\mu$ of the disorder
operator in the CFT~\cite{CFT}: $\eta^\star=2\Delta_\mu$.
We determine this scaling dimension numerically: To this end, we
extract the scaling dimensions from the spectrum of the transfer
operator $\mathbb{T}_\tau^\tau$ and classify them into different
topological sectors, shown in Fig.~\ref{DFib_CFT_spec} (b); by considering
the form factors of the
correlation function, we obtain that $\Delta_\mu=2/15$, and thus
$\eta^\star=4/15$, as discussed in Appendix \ref{corr_form_factor}.
Interestingly, we find that the two critical exponents $\eta$ and $\eta^\star$ are equal.
This is actually a consequence of the duality of the Potts model.
 It can be numerically observed that under the duality transformation, the
following relations hold to numerical accuracy:
\begin{eqnarray}\label{duality_spec}
 \lambda_{\bm{1},j}^{\bm{1}}(K) &=&\lambda_{\bm{1},j}^{\bm{1}}(K^\star),
\quad \lambda_{\bm{b},j}^{\bm{b}}(K)
=\lambda_{\bm{b},j}^{\bm{b}}(K^\star),\notag\\ \label{duality}
\lambda_{\bm{1},j}^{\bm{b}}(K) &
=&\lambda_{\bm{b},j}^{\bm{1}}(K)=\lambda_{\bm{\tau},j}^{\bm{\tau}}(K^\star)
=\lambda_{\bm{\bar{\tau}},j}^{\bm{\bar{\tau}}}(K^\star), \end{eqnarray}
  where $\lambda_{\bm{\alpha},j}^{\bm{\beta}}$ is the $(j+1)$-th dominant
eigenvalue belonging to the topological sector
$\langle\bm{\alpha}|\bm{\beta}\rangle$, where we rescale all $\lambda$ such that
$\lambda_{\bm{1},0}^{\bm{1}}=1$.  Therefore, the eigenvalues of the sectors
$\langle\bm{1}|\bm{b}\rangle$, $\langle\bm{b}|\bm{1}\rangle$,
$\langle\bm{\tau}|\bm{\tau}\rangle$ and
$\langle\bm{\bar{\tau}}|\bm{\bar{\tau}}\rangle$ are equal at the critical
point, as shown in Figs. \ref{DFib_CFT_spec} (a) and (b).
Importantly, these relations are a manifestation of a duality between the
topological sectors characterizing condensation and deconfinement,
respectively.  It is thus natural to conjecture that also the condensate
and the confinement fractions are dual to each other and their critical
exponents are the same: $\beta^{\star}=\beta=1/10$ and
$\nu^{\star}=\nu=3/4$.

\subsection{Numerical results}
\begin{figure*}
  \centering
  \includegraphics[width=18cm]{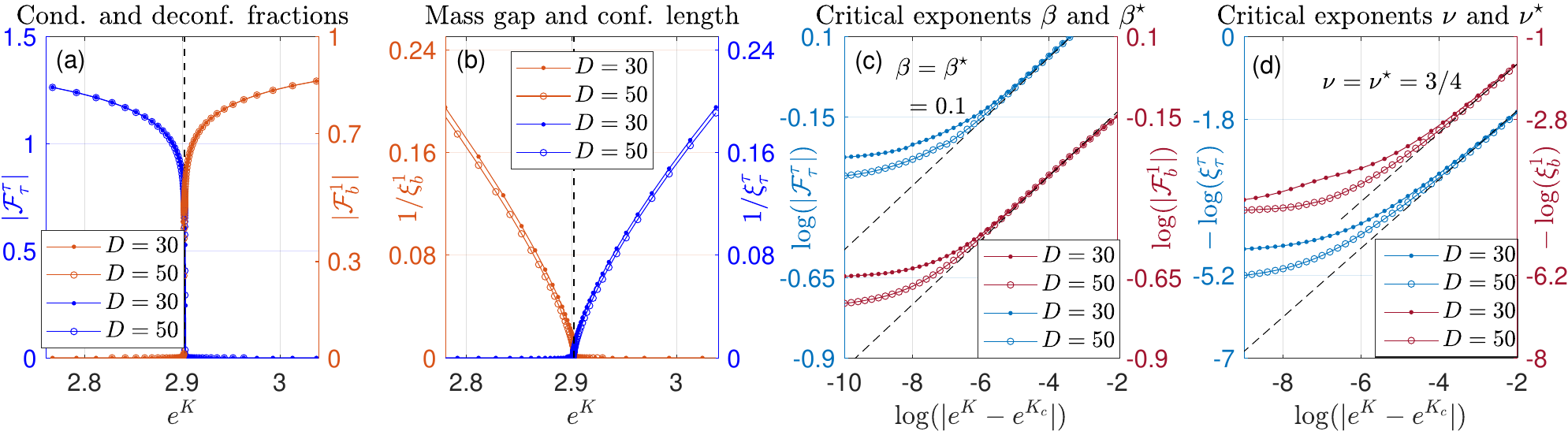}
  \caption{Numerical results for the DFib case. (a) The condensate fraction $|\mathcal{F}_{\bm{b}}^{\bm{1}}|$ and
  the deconfinement fraction $|\mathcal{F}_{\bm{\tau}}^{\bm{\tau}}|$ for different bond dimensions $D$.
  The dashed line indicates the exact position of the critical point.
  (b) The anyon ``mass gap'' $1/\xi_{\bm{b}}^{\bm{1}}$ and the inverse of the confinement length
  $1/\xi_{\bm{\tau}}^{\bm{\tau}}$.
  (c) Scaling of $|\mathcal{F}_{\bm{b}}^{\bm{1}}|$ and $|\mathcal{F}_{\bm{\tau}}^{\bm{\tau}}|$
  in the vicinity of the exact critical point; the slope of the dashed lines is the
    analytical prediction 1/10.
  (d) Scaling of $1/\xi_{\bm{\tau}}^{\bm{\tau}}$ and $1/\xi_{\bm{b}}^{\bm{1}}$
  in the vicinity of the exact critical point; the slope of the dashed lines
is the analytical predicition 3/4. }\label{DFib}
\end{figure*}
The predictions and conjectures above can be verified numerically.
The fractions and correlation lengths as well as their critical exponents
can be evaluated efficiently using well-established tensor network algorithms,
such as VUMPS\cite{VUMPS_2019,Verstraete_corner_2018} and CTMRG \cite{corboz_corner_2014}.
In Appendix \ref{Numerical_method}, the basic ideas of these tensor network algorithms are explained.
Since the transfer operator is Hermitian in the Fibonacci case, we use the
VUMPS method in the following.

Fig. \ref{DFib}  (a) shows the condensate fraction $|\mathcal{F}_{\bm{b}}^{\bm{1}}|$
and deconfinement fraction  $|\mathcal{F}_{\bm{\tau}}^{\bm{\tau}}|$ calculated using
different bond dimensions $D$ of the boundary matrix product states.
The position of the critical point
obtained from the condensate and deconfinement
fractions perfectly matches the exact value
$e^{K_c}\approx2.9021$.
Fig. \ref{DFib} (c) displays the scaling of $|\mathcal{F}_{\bm{b}}^{\bm{1}}|$
and $|\mathcal{F}_{\bm{\tau}}^{\bm{\tau}}|$  near
the exact critical point.
In the regime where the data is converged in $D$, the slopes (in a
log-log-plot) agree very well with
the critical exponents $\beta=\beta^\star=1/10$ predicted from the CFT.

Fig. \ref{DFib}  (b) shows the ``mass gap'' $1/\xi_{\bm{b}}^{\bm{1}}$ and
the inverse of the confinement length $1/\xi_{\bm{\tau}}^{\bm{\tau}}$.
The position of the critical point is consistent with the known value of
$e^{K_c}$.  Analyzing the scaling of $1/\xi_{\bm{b}}^{\bm{1}}$ and
$1/\xi_{\bm{\tau}}^{\bm{\tau}}$ close to the critical point in
Fig.~\ref{DFib} (d), we find that the observed slopes are consistent with
the critical exponents $\nu=\nu^\star=3/4$ predicted from CFT.

\section{Yang-Lee string-net}

\subsection{Deformed DLY string-net wavefunction}
\begin{figure*}
  \centering
  \includegraphics[width=16cm]{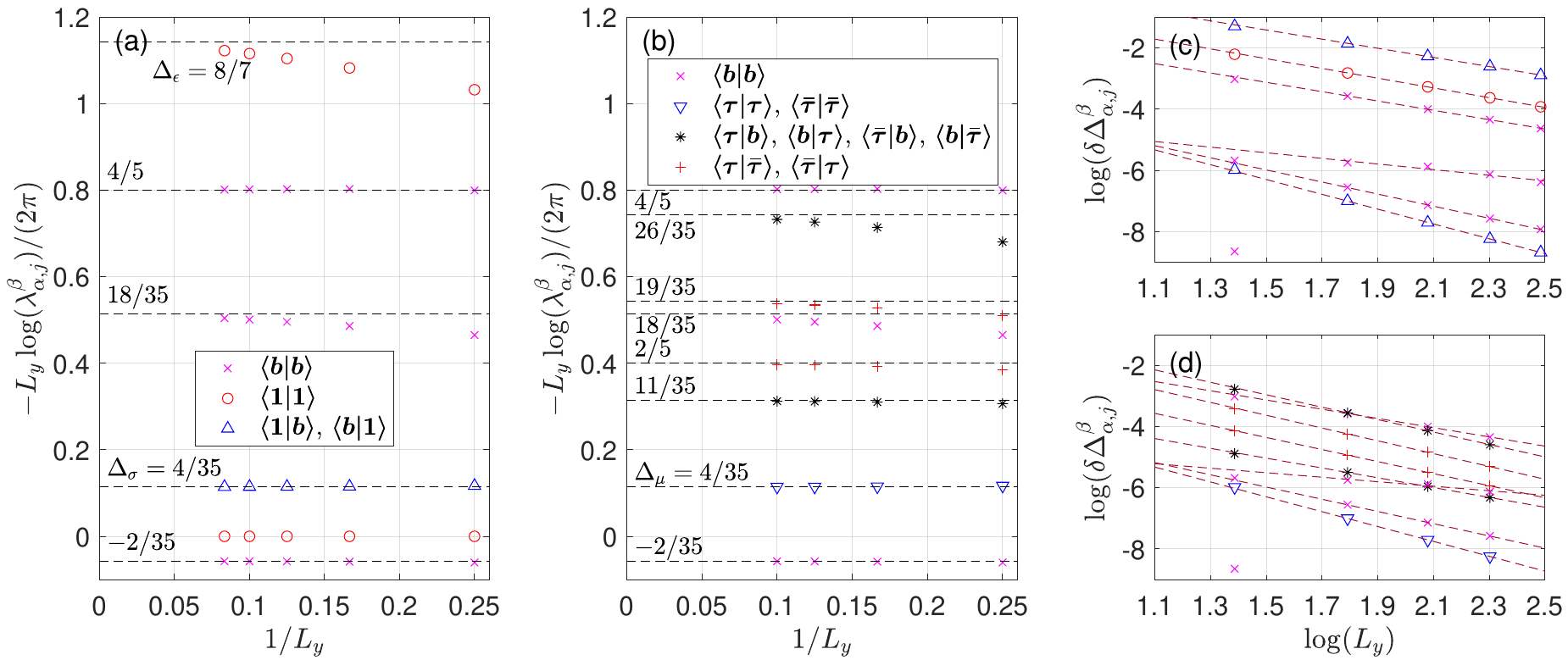}
  \caption{Numerically extracted scaling dimensions of primary fields
in the DYL case. (a)-(b) Scaling dimensions
$-L_y\log(\lambda_{\bm{\alpha},j}^{\bm{\beta}})/2\pi$ extracted numerically
from $\mathbb{T}_1^1$ (left) and $\mathbb{T}_{\tau}^{\tau}$
(right)
on cylinders
with circumference $L_y$, where
$\lambda_{\bm{\alpha},j}^{\bm{\beta}}$ are rescaled eigenvalues,
classified into different topological sectors $\langle\bm{\alpha}|\bm{\beta}\rangle$.
The dashed lines are the CFT predictions.  $\Delta_\sigma$, $\Delta_\epsilon$ and $\Delta_\mu$
label the scaling dimensions of order parameters, energy, and disorder
operators. (c)-(d) The finite-size corrections
$\delta\Delta_{\bm{\alpha},j}^{\bm{\beta}}(L_y)=|-L_y\log(\lambda_{\bm{\alpha},j}^{\bm{\beta}})/2\pi-\Delta_{\bm{\alpha},j}^{\bm{\beta}}|$
of the scaling dimensions extracted from $\mathbb{T}_{1}^{1}$ (above) and
$\mathbb{T}_{\tau}^{\tau}$ (below) vanish algebraically with $L_y$, where
$\Delta_{\bm{\alpha},j}^{\bm{\beta}}$ are the exact values predicted from
CFT.}\label{DYL_CFT_spec} \end{figure*}
In analogy to the deformed DFib string-net wavefunction, the deformed
Yang-Lee string-net wavefunction can also be obtained by acting with
deformation operators on
the $\{i_1\}$ and $\{i_2\}$ spins of a
ground state wavefunction
$|\Psi_\text{DYL}\rangle_R$ of the DYL string-net model:
\begin{equation}\label{deformed_DYL}
|\Psi(K)\rangle_R=\prod_{i_1i_2}\exp\left[K(\sigma^z_{i_1}+\sigma^z_{i_2})/4\right]|\Psi_\text{DYL}\rangle_R.
\end{equation}

The parent Hamiltonian for the fixed point model at $K=0$ is complex
symmetric, $H=H^T$, and has a real spectrum~\cite{Galois_conjugate_2012},
and thus, the left
ground state eigenvector is the transpose of the right one,
$_L\langle\Psi(0)\vert=\big(\vert\Psi(0)\rangle_R\big)^T$; we henceforth
distinguish them by subscripts $L$ and $R$. While it is not
clear how to modify the Hamiltonian for the DYL model such as to have
$\vert\Psi(K)\rangle_R$ as its ground state [a modification analogous to
Eq.~\eqref{eq:ham-deformed-dfib} does not necessarily have positive
spectrum, as $H(0)$ is
not positive semi-definite], we anticipate that a suitable
parent Hamiltonian should keep the property that
$_L\langle\Psi(K)\vert=\big(\vert\Psi(K)\rangle_R\big)^T$, which we assume
henceforth.
Due to the non-Hermitian nature of the system (where the normalization
condition imposes that left and right
eigenvectors are biorthogonal), it is natural to consider the overlap
$_L\langle\Psi(K)|\Psi(K)\rangle_R$, rather than the normalization of
$|\Psi(K)\rangle_R$, when mapping the system to a statmech model, and
correspondingly
when constructing condensation and deconfinement order parameters by
inserting MPOs, and we will do so in the following.

Since the DYL string-net wavefunction can be obtained from the DFib string-net
wavefunction by substituting $\phi^\prime=-1/\phi$ for $\phi$, it is
natural to expect that
for the
deformed DYL wavefunction,
$_L\langle\Psi(K)|\Psi(K)\rangle_R$
can be mapped to the partition function of the
$(\phi^\prime+2)$-state Potts model. In Appendix \ref{Map_to_RSOS}, we
prove -- using the formulation in terms of tensor networks --
that $_L\langle\Psi(K)|\Psi(K)\rangle_R$ for the deformed DYL wavefunction
is exactly equivalent to the partition function of a non-unitary RSOS
model
associated with the $D_6$ Dynkin diagram.  From there, we can again
conclude
from the Potts--RSOS equivalence \cite{he_2020_geometrical} that
this nonunitary RSOS model is in turn equivalent to the square lattice
$(\phi^\prime+2)$-state Potts model. Thus, we have
\begin{equation}\label{quantum_classical_mapping_DYL}
{}_L\langle\Psi(K)|\Psi(K)\rangle_R\propto\mathcal{Z}_{\text{Potts}}(\phi^\prime+2,K).
\end{equation} According to Eq. \eqref{central_charge}, the critical
point of the $(\phi^{\prime}+2)$-state Potts model is located at
$e^{K_c}=1+\sqrt{2+\phi^{\prime}}\approx2.1756$ and is described by a
non-unitary minimal CFT with a central charge $c=8/35$.

\subsection{Condensate and deconfinement fractions}
\begin{figure*}
  \centering
  \includegraphics[width=18cm]{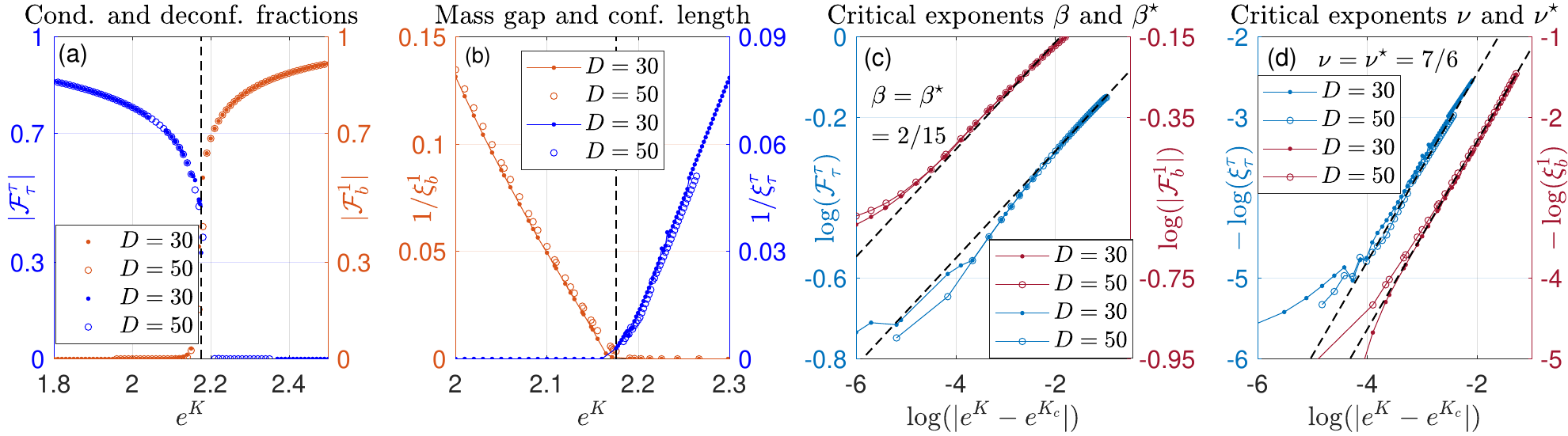}
  \caption{Numerical results for the DYL case. (a) Condensate
fraction $|\mathcal{F}_{\bm{b}}^{\bm{1}}|$ and deconfinement fraction
$|\mathcal{F}_{\bm{\tau}}^{\bm{\tau}}|$ for different bond dimensions $D$.
the dashed line indicates the exact position of the critical point.  (b)
Anyon ``mass gap'' $1/\xi_{\bm{b}}^{\bm{1}}$ and inverse of the
confinement length $1/\xi_{\bm{\tau}}^{\bm{\tau}}$.  (c) Scaling of
the $|\mathcal{F}_{\bm{b}}^{\bm{1}}|$ and
$|\mathcal{F}_{\bm{\tau}}^{\bm{\tau}}|$ in the vicinity of the exact
critical point.
The slope of the dashed lines is the CFT prediction 2/15.  (d) Scaling of
$1/\xi_{\bm{\tau}}^{\bm{\tau}}$ and $1/\xi_{\bm{b}}^{\bm{1}}$ in the
vicinity of the exact critical point. The slope of the dashed lines is the
CFT prediction 7/6.
}\label{Fig_Yang_Lee}
\end{figure*}
The condensate fraction for the Yang-Lee case is:
\begin{equation}
 \mathcal{F}^{\bm{1}}_{\bm{b}}= \frac {{}_L\langle \bm{1}|\dot{\bm{b}}_i\rangle_{R}}{{}_L\langle \bm{1}|\bm{1}\rangle_R},
\end{equation}
where $|\bm{1}\rangle$ is a MES and $|\dot{\bm{b}}_i\rangle_{R}$ is
obtained by creating a $\bm{b}$ excitation on top of $|\bm{1}\rangle$.
Since the left and right eigenvectors satisfy bi-orthogonality, the
analysis is analogous to the one in the Fibonacci case.  The condensate
fraction $\mathcal{F}^{\bm{1}}_{\bm{b}}$ is zero in the topological phase
and non-zero in the non-topological phase.  Yet again,
$\mathcal{F}^{\bm{1}}_{\bm{b}}$ and $\mathcal{F}^{\bm{b}}_{\bm{1}}$ can be
mapped to the expectation values of the local order parameters of the
RSOS model, as proved in Appendix \ref{Map_to_RSOS}.

The deconfinement fraction is
\begin{equation}
  \mathcal{F}_{\bm{\tau}}^{\bm{\tau}}=\frac{{}_L\langle\dot{\bm{\tau}}_i|\dot{\bm{\tau}}_i\rangle_R}{{}_L\langle\bm{1}|\bm{1}\rangle_R},
\end{equation}
where the definition of $|\dot{\bm{\tau}}_i\rangle_R$ is the same as
in the DFib case.  Because the $\bm{\tau}$ anyons are chiral,
the left eigenvector ${}_L\langle\dot{\bm{\tau}}_i|$ is not simply the transpose
of the right eigenvector $|\dot{\bm{\tau}}_i\rangle_R$:
Considering that an excitation carried by a left eigenvector should have
a chirality opposite to that of
the excitation carried by the corresponding right eigenvector,
 the proper definition of ${}_L\langle\bm{\tau}_i|$ is to insert an end tensor
\begin{equation}
 E_{\bm{\tau}}(d_i,F_{tsu}^{ijk},(R^{ij}_k)^\star)=E_{\bm{\bar{\tau}}}(d_i,F_{tsu}^{ijk},R^{ij}_k)
\end{equation}
with an infinite long non-trivial MPO string attached to it into the PEPS ${}_L\langle\bm{1}|$.
Again, the deconfinement fraction can be considered as a non-local
disorder parameter for the $(\phi^{\prime}+2)$-state Potts model.

The correlation functions $\mathcal{C}_{\bm{b}}^{\bm{1}}$ and
$\mathcal{C}_{\bm{\tau}}^{\bm{\tau}}$ in the DYL case are similar to those
in the DFib case: $\mathcal{C}_{\bm{b}}^{\bm{1}}$ ($\mathcal{C}_{\bm{\tau}}^{\bm{\tau}}$) decays exponentially in the gapped topological (non-topological) phase:
\begin{eqnarray}
  \mathcal{C}_{\bm{b}}^{\bm{1}}(|i-j|)&=&\frac{{}_L\langle\bm{1}|\dot{\bm{b}}_i\dot{\bm{b}}_j\rangle_R}{{}_L\langle\bm{1}|\bm{1}\rangle_R}\propto \exp{\left(-\frac{|i-j|}{\xi^{\bm{1}}_{\bm{b}}}\right)},\notag\\
  \mathcal{C}_{\bm{\tau}}^{\bm{\tau}}(|i-j|)&=&\frac {{}_L\langle \dot{\bm{\tau}}_i\dot{\bm{\tau}}_j|\dot{\bm{\tau}}_i\dot{\bm{\tau}}_j\rangle_R}{{}_L\langle \bm{1}|\bm{1}\rangle_R}\propto \exp{\left(-\frac{|i-j|}{\xi_{\bm{\tau}}^{\bm{\tau}}}\right)},
\end{eqnarray}
where $|\dot{\bm{b}}_i\dot{\bm{b}}_j\rangle_R$ and $|\dot{\bm{\tau}}_i\dot{\bm{\tau}}_j\rangle_R$
are obtained by creating a pair of anyons on top of $|\bm{1}\rangle_R$, and the anyons in ${}_L\langle \dot{\bm{\tau}}_i\dot{\bm{\tau}}_j|$
carry the opposite chirality.
Again, the scaling of these quantities close to criticality is given by
scaling exponents $\beta$, $\nu$, $\eta$, and
$\beta^\star$, $\nu^\star$, $\eta^\star$, as defined in
Eqs.~\eqref{exponents_near_K_c} and  \eqref{exponents_at_K_c}.

To predict the critical exponents from CFT, we proceed as the
Fibonacci case. $\Delta_\sigma$ and $\Delta_\epsilon$ can be determined by
known results for the corresponding RSOS and Potts models, as shown in Appendix
\ref{Map_to_RSOS}. Again, $\Delta_\mu$ can only be identified
numerically from the spectrum of the transfer operator in the
respective sector; the corresponding numerical results are discussed in  Fig.~\ref{DYL_CFT_spec}(b) and Appendix
\ref{corr_form_factor}.
Note that there is a negative scaling dimension $-2/35$ in
the topological sector $\langle\bm{b}|\bm{b}\rangle$,
signifying the non-unitarity of the CFT. In summary, we obtain
\begin{equation}
\Delta_\sigma=\Delta_\mu=4/35,\quad \Delta_\epsilon=8/7.  \end{equation}
Notice that the duality \eqref{duality_spec} between different
 topological sectors is still satisfied for the DYL case.
 Assuming that the scaling relations $\Delta_\sigma=\eta/2,\Delta_\epsilon=2-1/\nu, \beta
=\eta\nu/2$ are still valid for this non-Hermitian model,
 the critical exponents are obtained as:
\begin{equation}
  \eta=8/35,\quad \nu=7/6,\quad \beta=2/15.
\end{equation}
From duality, we expect the dual critical exponents $\eta^{\star}$, $\mu^{\star}$ and $\beta^{\star}$
to be equal to $\eta$, $\mu$ and $\beta$, respectively.

\subsection{Numerical results}

The predictions and expectations above can be verified numerically
by computing the topological order and disorder parameters.
Because in the DYL case the transfer operator is non-Hermitian, the
VUMPS method cannot be reliably used, and we resort to CTMRG instead.
Fig.~\ref{Fig_Yang_Lee}(a) shows the condensate fraction
$|\mathcal{F}_{\bm{b}}^{\bm{1}}|$ and the deconfinement fractions
$|\mathcal{F}_{\bm{\tau}}^{\bm{\tau}}|$ calculated using different bond
dimensions $D$ of the CTM environments.  The position of the critical
point implied by the fractions matches the exact value
$e^{K_c}=1+\sqrt{2+\phi^{\prime}}\approx2.1756$.  Fig. \ref{Fig_Yang_Lee} (c)
displays the scaling of condensate and deconfinement fractions, and the
slopes of the data which are converged in $D$ are in good agreement with
critical exponents $\beta=\beta^\star=2/15$ predicted from CFT.  Fig.
\ref{Fig_Yang_Lee}  (b) shows the ``mass gap'' and the inverse of the confinement
length, and Fig. \ref{Fig_Yang_Lee} (d) shows their scaling close to criticality, which
is again in good agreement with the analytically derived critical
exponents $\nu=\nu^\star=7/6$.

\section{Conclusion and discussion}

In this work, we have generalized order parameters for condensation and
deconfinement of anyons to the case of non-Abelian topological models,
including no-Hermitian ones.  In particular, we have focused on the DFib
string-net model and its Galois conjugate, the DYL string-net model. We
proved that the normalization of the string tension deformed DFib  and DYL
string-net states are equivalent to partition functions of $D_6$ RSOS
models, and based on this we proved the equivalence of the condensate fraction
to the order parameter of the RSOS models. We have used this equivalence
to predict critical exponents related to the condensation and
deconfinement order parameters, and we have confirmed these predictions by
numerical studies.

Our approach can be straightforwardly applied to other non-Abelian
string-net models. An interesting additional aspect beyond condensation
and confinement of anyons, which arises in topological phase transitions
in more complex models, is the fact that one non-Abelian anyon can
split into different kinds of anyons at the transition~\cite{burnell_anyon_2018,D_4_PEPS}.
It is an interesting question whether and how to construct an order parameter
that directly detects this splitting.  A different question is how to
generalize our approach to variationally opimized PEPS obtained from
tuning a Hamiltonian, rather than from explicitly constructed wavefunction
families. To this end, one could use either a constrained optimization on
the manifold of MPO-symmetric tensors~\cite{Iqbal_2020}, or 
methods developed to extract the MPO symmetry a posteriori from an
optimized PEPS tensor~\cite{Anna_2020}. It is an interesting open
question to construct and investigate condensation and deconfinement order
parameters in such a scenario.

\begin{acknowledgements}

W.-T.~Xu would like to gratefully acknowledge the early help of Qi Zhang and Hai-Jun Liao on
VUMPS and CTMRG programs. This work has been supported by the European
Research Council (ERC) under the European Union's Horizon 2020 research
and innovation programme through the ERC-CoG SEQUAM (Grant Agreement
No.~863476). The computational results presented have been achieved in
part using the Vienna Scientific Cluster (VSC).

\end{acknowledgements}

\appendix

\section{Definitions of all tensors}\label{Definitions_of_all_tensors}
All tensors can be defined using $d_i,N_{ij}^k, F^{ijk}_{tsu}$ and $R^{ij}_k$. The $d$ tensor has already been defined in the main text. The non-zero entries of the $N$ tensor are
$N_{11}^1=N_{1\tau}^\tau=N_{\tau1}^\tau=N_{\tau\tau}^1=N_{\tau\tau}^\tau=1$.
The non-zero entries of the $F$ tensor are determined by $N_{ij}^k$,
i.e., $F^{ijk}_{tsu}\neq0$ if $N_{ij}^kN_{ts}^kN_{is}^uN_{jt}^u=1$. For
the DFib and DYL cases, the nontrivial entries are given
by Eqs. \eqref{DFib_F_symbol} and \eqref{DYL_F_symbol}, separately, and other non-zero entries are $1$. From the $F$ tensor, it is convenient to define the $G$ tensor:
\begin{equation}
  G^{ijk}_{\alpha\beta\gamma}=F^{ijk}_{\alpha\beta\gamma}/\sqrt{d_kd_\gamma}=\includegraphics[width=1.5cm,valign=c]{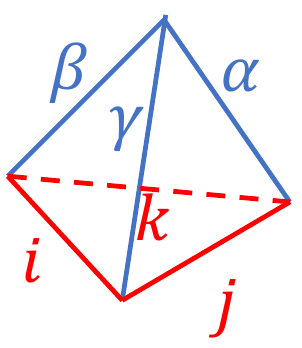}.
\end{equation}
The $G$ tensor has tetrahedral symmetry so we can also represent it using a tetrahedron.  A triple-line local tensor generating the PEPS of a string-net wavefunction can be expressed as\cite{GuTensorNetwork_2009,buerschaper_explicit_2009}:
\begin{equation}\label{string-net_tensor}
 \includegraphics[width=2cm,valign=c]{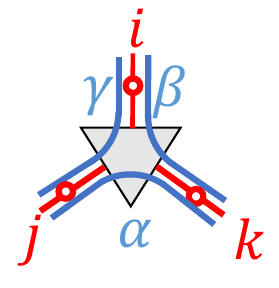}=(d_{i}d_{j}d_{k})^\frac{1}{4}G^{ijk}_{\alpha\beta\gamma}
\end{equation}
where the open circles represent the physical degrees of freedom and the lines are virtual degrees of
freedom. Each line represents a $\delta$ tensor, i.e., the entries are non-zero iff all of its indices are equal.
 In the PEPS representation of the string-net wavefunctions,
 there is a convention that the contraction of the degrees of freedom
  is a sum weighted by the quantum dimenisons, so we should assign the weight $d_\alpha^{\frac{1}{6}}, d_\beta^{\frac{1}{6}}$ and $d_\gamma^{\frac{1}{6}}$ to $\alpha$, $\beta$ and $\gamma$ indices of the tensor in \eqref{string-net_tensor}, but we omit them after and in the next for convenience. The local tensor on the square lattice is obtained
  by contracting the above two tensors:
\begin{equation}
 \includegraphics[width=3cm,valign=c]{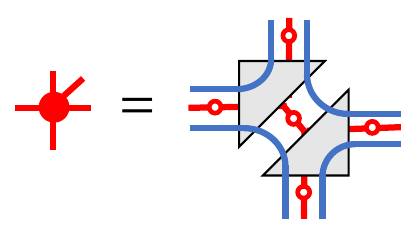}
\end{equation}

The local tensor of the horizontal MPO $O_n$ is\cite{MPO_algebra_2017}:
\begin{equation}\label{MPO_tensor}
 \includegraphics[width=3cm,valign=c]{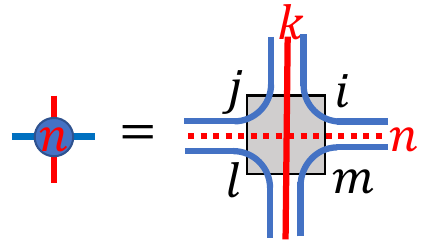}=G_{lmn}^{ijk},
\end{equation}
where $n=1$ or $\tau$ is a fixed index. In the abbreviated graphs, the red (blue) lines represent
the triple-line (double-line) in the original graphs, and the horizontal MPO can be generated by the tensor:
\begin{equation}
 \includegraphics[width=4.5cm,valign=c]{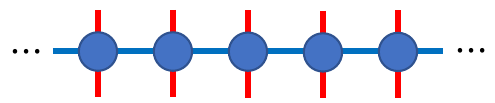}.
\end{equation}
Furthermore, by defining the tensor with the fixed indices $i,j$ and $k$:
\begin{equation}
 \includegraphics[width=4cm,valign=c]{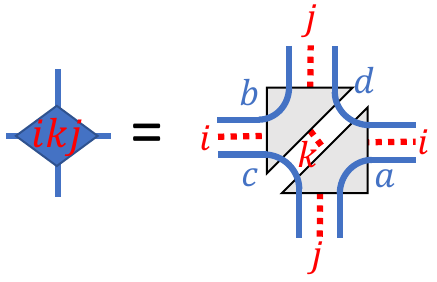}=G^{ijk}_{dcb}G^{jki}_{dac},
\end{equation}
a vertical MPO $O_{ikj}$ can be generated together with the tensor (it should be rotated by $\pi/2$) in Eq. \eqref{MPO_tensor}:
\begin{equation}
 O_{ikj}=\includegraphics[width=1cm,valign=c]{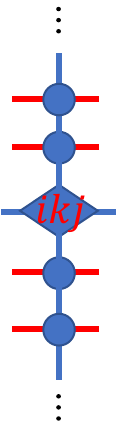}.
\end{equation}
Notice that the up and down legs are connected periodically.
These vertical MPOs form a basis of the tube algebra.
The following linear combinations of the vertical MPOs are the central idempotents of the tube algebra:
\begin{equation}
P_{\bm{\alpha}}=\frac{d_a d_b }{d_1^2+d_\tau^2}\sum_{ikj}d_j d_kC_{ikj}(a,b)O_{ikj},\\
\end{equation}
where indices $(a,b)$ are determined by $\bm{\alpha}$:
\begin{equation}
  \bm{1}=(1,1), \quad   \bm{\tau}=(\tau,1),\quad   \bm{\bar{\tau}}=(1,\tau),\quad   \bm{b}=(\tau,\tau),
\end{equation}
and
\begin{equation}
  C_{ikj}(a,b)=\sum_{\gamma\delta} d_\gamma d_\delta R^{aj}_{\gamma}  R^{j b}_{\delta}G^{i\delta\gamma}_{j a b}G^{k a \delta}_{b j i}G^{i k j}_{a \gamma \delta}.
\end{equation}
The explicit expressions of the idempotents for the DFib and DYL cases can be found in Refs. \cite{Schotte_2020_Fibonacci} and \cite{Galois_strange_correlator}, respectively.
The non-zero entries of $R$ tensor are
\begin{eqnarray}
  R_1^{11}&=&R_{\tau}^{1\tau}=R_{\tau}^{\tau1}=1,\notag\\
   R_1^{\tau\tau}&=&\begin{cases}
e^{4\pi i/5}, & \text{DFib }\\
e^{2\pi i/5}, & \text{DYL }
\end{cases},\quad R_\tau^{\tau\tau}=\begin{cases}
e^{-3\pi i/5}, & \text{DFib }\\
e^{\pi i/5}, & \text{DYL }
\end{cases}.\notag
\end{eqnarray}
Since $C_{\tau kj}(1,1)=C_{1 kj}(1,\tau)=C_{1 kj}(\tau,1)=0$, we have
\begin{equation}\label{Idempotents}
  P_{\bm{1}}=P_{\bm{1}1},\quad P_{\bm{\tau}}=P_{\bm{\tau}\tau},\quad P_{\bm{\bar{\tau}}}=P_{\bm{\bar{\tau}}\tau},\quad P_{\bm{b}}=P_{\bm{b}1}\oplus P_{\bm{b}\tau}.
\end{equation}

The end tensor carrying anyonic excitations is defined as\cite{Schotte_2020_Fibonacci}
%\begin{widetext}
\begin{eqnarray}\label{end_tensor}
&& E_{\bm{\alpha} k}=\includegraphics[width=4cm,valign=c]{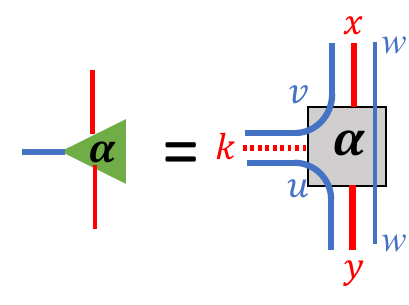}\notag\\
 &&=d_a^{\frac{1}{4}}d_b^{\frac{1}{4}}d_x^{\frac{1}{4}}d_y^{\frac{1}{4}}d_k\sum_{\beta}d_\beta C_{k\beta w}(a,b)G^{u\beta x}_{w vk}G_{x\beta k}^{wyu}.\notag\\
\end{eqnarray}

\section{Mapping the deformed PEPS to the RSOS models}\label{Map_to_RSOS}
 The norms of the string-net wavefunctions can be exactly mapped to the partition function of the RSOS models, from which we know the positions of critical points and the CFTs describing the critical points. Furthermore, we can also find that the condensate fractions are exactly the same as the expectation values of the RSOS order parameters, from which one can find the critical exponents related to the condensate fractions.

 At first we consider a local double tensor for the norm of the DFib or DYL PEPS without the deformation:
 \begin{equation}\label{double_tensor}
\sum_{k}d_k\includegraphics[width=2.5cm,valign=c]{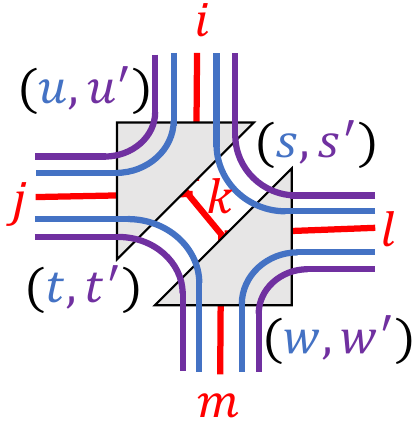}
=\sum_{k}d_kG^{ijk}_{tsu}G^{klm}_{wts}G^{ijk}_{t^\prime s^\prime u^\prime}G^{klm}_{w^\prime t^\prime s^\prime}.
\end{equation}
Since the $F$-move gives rise to the following relation:
 \begin{equation}\label{square_symmetry}
\sum_{k}d_k\includegraphics[width=2.5cm,valign=c]{double_tensor_1.pdf}
=\sum_{n}d_{n}\includegraphics[width=2.5cm,valign=c]{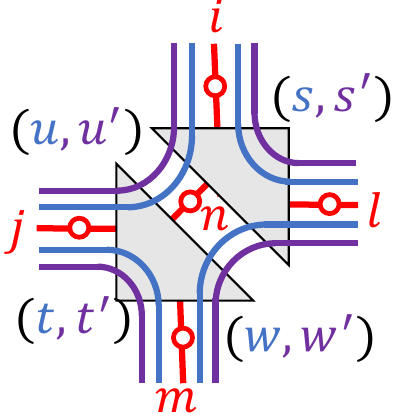},
\end{equation}
 the tensor in \eqref{double_tensor} actually respects the symmetry described the dihedral group $D_4$.
In addition, because the $G$ tensor has the tetrahedral symmetry, the relation \eqref{square_symmetry} can also be represented by the tetrahedrons:
 \begin{equation}
\sum_{k}d_k\includegraphics[width=2.2cm,valign=c]{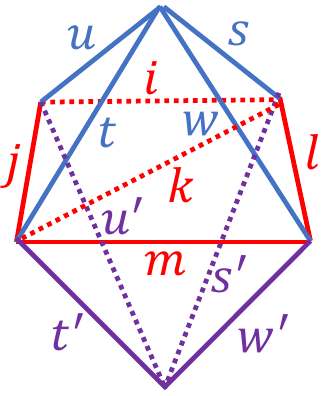}
=\sum_{n}d_{n}\includegraphics[width=2.2cm,valign=c]{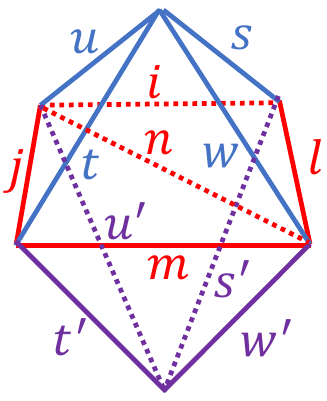},
\end{equation}
where each $G$ tensor is represented by a tetrahedron.

In this tetrahedral representation, one can easily find the following relation:
 \begin{equation}
\sum_{k}d_k\includegraphics[width=2.2cm,valign=c]{Tetrahedra_1.pdf}
=\sum_{x}d_{x}\includegraphics[width=2.2cm,valign=c]{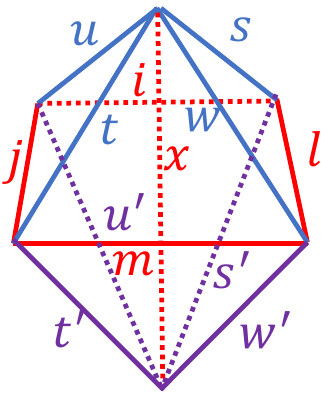}.
\end{equation}
Namely
 \begin{eqnarray}
&&\sum_{k}d_kG^{ijk}_{tsu}G^{klm}_{wts}G^{ijk}_{t^\prime s^\prime u^\prime}G^{klm}_{w^\prime t^\prime s^\prime}\notag\\
&=&\sum_{x}d_x G^{utj}_{t^\prime u^\prime x}G^{twm}_{w^\prime t^\prime x}G^{wsl}_{s^\prime w^\prime x}G^{sui}_{u^\prime s^\prime x}.
\end{eqnarray}
So a double tensor at a vertex of the lattice is decomposed into four $G$ tensors living seperately on the four edges:
 \begin{equation}\label{decomposition}
\sum_{k}d_k\includegraphics[width=2.5cm,valign=c]{double_tensor_1.pdf}
=\sum_{x}d_{x}\includegraphics[width=2.5cm,valign=c]{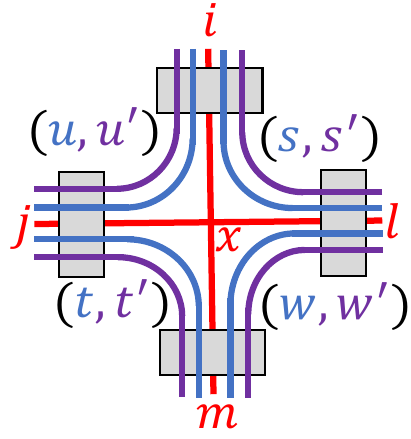}.
\end{equation}

\begin{figure}
  \centering
  \includegraphics[width=8.5cm]{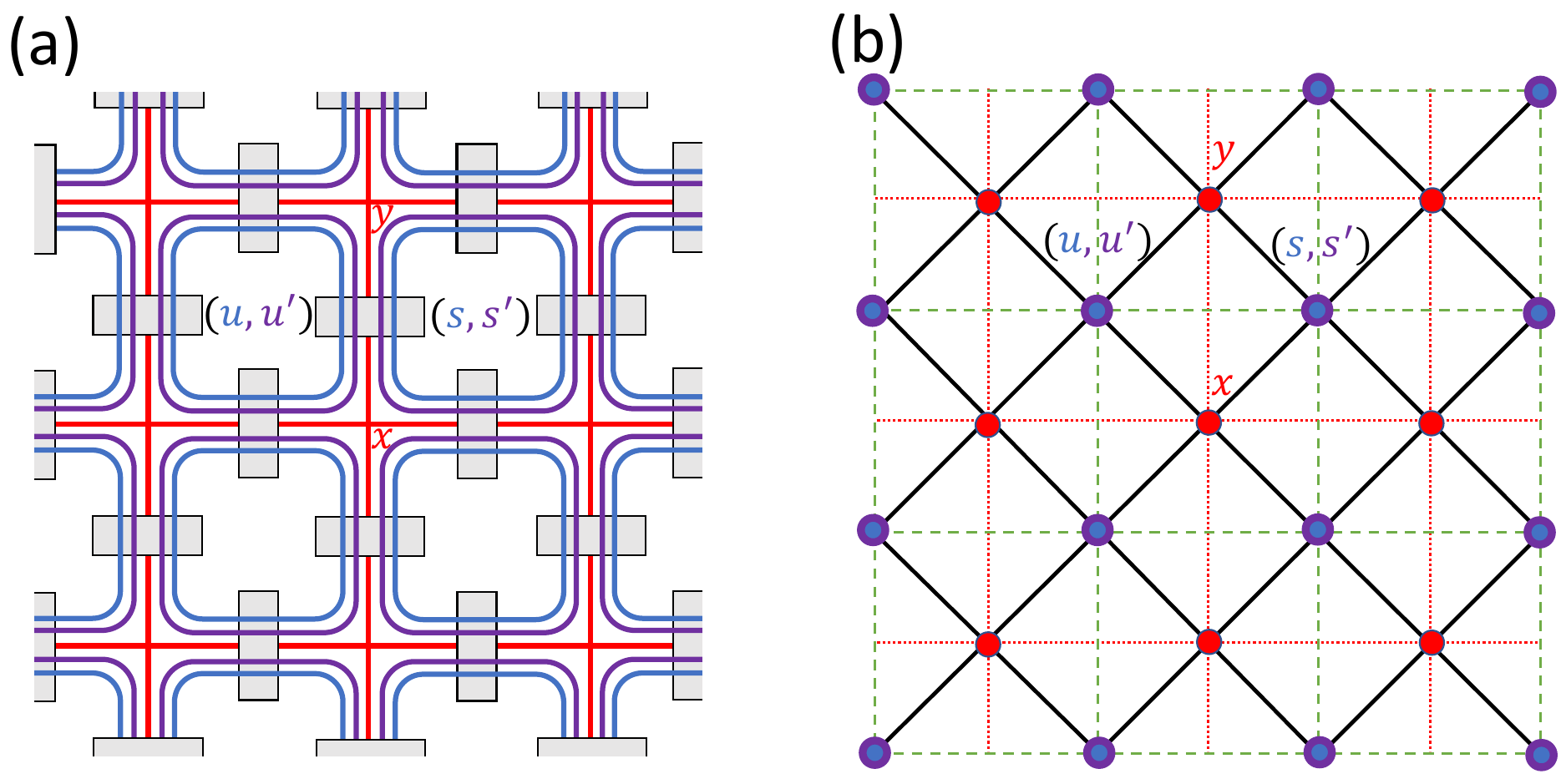}
  \caption{(a) A tensor network representation for the RSOS models. (b) The degrees of freedom of the RSOS models are represented by dots. The red/green/black lines form the primal/dual/medial lattice.}\label{RSOS}
\end{figure}

Since the two nearest neighboring original double tensors share a common edge, there are two $G$ tensors on each edge after the decomposition \eqref{decomposition}, and we can contract them together with the string tension deformation $T=(e^{K/2},e^{-K/2})$:
\begin{eqnarray}\label{edge_tensor}
   &&\sum_{i}d_iG^{sui}_{u^\prime s^\prime x}G^{usi}_{s^\prime u^\prime y}T_i
   = \sum_{i}d_iG^{sui}_{u^\prime s^\prime x}G^{sui}_{u^\prime s^\prime y}T_i\notag\\
   &=&e^{-K/2}\left(G^{su1}_{u^\prime s^\prime x}G^{su1}_{u^\prime s^\prime y}e^K+
   d_\tau G^{su\tau}_{u^\prime s^\prime x}G^{su\tau}_{u^\prime s^\prime y}\right) \notag\\
   &=&e^{-K/2}\left[G^{su1}_{u^\prime s^\prime x}G^{su1}_{u^\prime s^\prime y}(e^K-1)+\sum_{i}
   d_i G^{sui}_{u^\prime s^\prime x}G^{sui}_{u^\prime s^\prime y}\right]\notag\\
   &=&e^{-K/2}\left[\frac{\delta_{ss^\prime}\delta_{uu^\prime}}{d_sd_u}(e^K-1)+\frac{\delta_{xy}}{d_x}\right]N_{ss^\prime}^{x}N_{uu^\prime}^{y}\notag\\
   &=&\frac{\sqrt{q}}{e^{K/2}}\left(\frac{\delta_{ss^\prime}\delta_{uu^\prime}}{d_sd_u}\frac{e^K-1}{\sqrt{q}}+\frac{\delta_{xy}}{d_x\sqrt{q}}\right)N_{ss^\prime}^{x}N_{uu^\prime}^{y}.
\end{eqnarray}
As shown in Fig. \ref{RSOS} (a), the tensor in \eqref{edge_tensor} is represented by the gray rectangles, and in the following, we will explain that the tensor network in Fig. \ref{RSOS} (a) is merely one of the RSOS models.

 As displayed in Fig. \ref{RSOS} (b), the red lines in Fig. \ref{RSOS} (a) are the degrees of freedom living on the sites of the primal square lattice and they takes two values $1$ and $\tau$. The blue and purple lines in Fig. \ref{RSOS} (a) are the degrees of freedom living on the sites of the dual square lattice and they takes four values $(1,1)$, $(\tau,\tau)$, $(1,\tau)$ and $(\tau,1)$. All of them are degrees of freedom of the RSOS models, which live on the medial lattice shown in \ref{RSOS} (b). The nearest neighboring  degrees of freedom of the RSOS models are restricted by two $N$ tensors in the last line of Eq. \eqref{edge_tensor}. And this restriction can be represented by the following $D_6$ Dynkin diagram\cite{pasquier_Dynkin_RSOS,pasquier_1987_lattice,pasquier_1987,Fibonacci_ladder_2009}:
 \begin{equation}\label{Dynkin_diagram}
\includegraphics[width=4cm,valign=c]{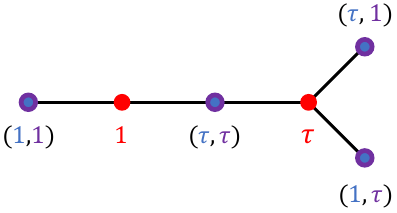}.
\end{equation}
This means that the two degrees of freedom of the RSOS models can be nearest neighboring if and only if they are adjacent in the Dynkin diagram. The nodes of the diagram are denoted by $h$, which are called heights of the RSOS models, and they take the six values $(1,1), 1,(\tau,\tau),\tau,(1,\tau),(\tau,1)$.  The Dynkin diagram imposes the restriction on the medial lattice and naturally divides the medial lattice into two sublattices, which are the primal lattice and the dual lattice. And the Dynkin diagram defines not only the configurations of the RSOS models but also their Boltzmann weights, as shown in the following.

At first we consider an adjacency matrix describing the Dynkin diagram \eqref{Dynkin_diagram}:
\begin{equation}
  \mathcal{A}=\left(
    \begin{array}{cccccc}
      0 & 1 & 0 & 0 & 0 & 0 \\
      1 & 0 & 1 & 0 & 0 & 0 \\
      0 & 1 & 0 & 1 & 0 & 0 \\
      0 & 0 & 1 & 0 & 1 & 1 \\
      0 & 0 & 0 & 1 & 0 & 0 \\
      0 & 0 & 0 & 1 & 0 & 0 \\
    \end{array}
  \right).
\end{equation}
Its eigenvalues are $2\cos(r\pi/p)$, where $p=10$ is the Coxeter number for the $D_6$ Dynkin diagram and $r=1,3,5,7,9$. The corresponding eigenvectors $S^{(r)}$ are
\begin{eqnarray}
    S^{(1)}&=&(1,\sqrt{\phi+2},\phi^2,\phi\sqrt{\phi+2},\phi,\phi)^T,\notag\\
    S^{(3)}&=&(1,\sqrt{\phi^\prime+2},\phi^{\prime2},\phi^\prime\sqrt{\phi^\prime+2},\phi^\prime,\phi^\prime)^T,\notag\\
    S^{(\bar{5})}&=&(1,0,-1,0,\phi,\phi^\prime)^T,\notag\\
    S^{(5)}&=&(1,0,-1,0,\phi^\prime,\phi)^T,\notag\\
    S^{(7)}&=&(1,-\sqrt{\phi^\prime+2},\phi^{\prime2},-\phi^\prime\sqrt{\phi^\prime+2},\phi^\prime,\phi^\prime)^T,\notag\\
    S^{(9)}&=&(1,-\sqrt{\phi+2},\phi^2,-\phi\sqrt{\phi+2},\phi,\phi)^T.\notag\\
\end{eqnarray}
The eigenvalue $0$ is two-fold degenerate, so we distinguish the corresponding two eigenvectors $S^{(5)}$ and $S^{(\bar{5})}$.

There are two different RSOS models. One is defined using $S=S^{(1)}$, and the other one is defined using $S=S^{(3)}$. The partition functions of the RSOS models can be expressed in terms of the adjacency matrix $\mathcal{A}$ and $S$:
\begin{equation}\label{RSOS_partition_function}
  \mathcal{Z}_\text{RSOS}=\sum_{\{h_i\}}\prod_{\langle h_ih_j\rangle}\mathcal{A}_{h_i,h_j}
  \prod_{i}S_{h_i}\prod_{\diamond}W_\diamond,
\end{equation}
where
\begin{eqnarray}\label{RSOS_Boltzmann_weight}
  W_\diamond &=& W(\includegraphics[width=1.5cm,valign=c]{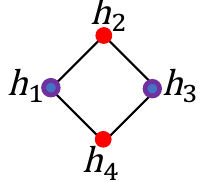})=\frac{\delta_{h_1h_3}}{S_{h_1}}\frac{e^K-1}{\sqrt{q}}+\frac{\delta_{h_2h_4}}{S_{h_2}} \notag\\
 \text{or} && W(\includegraphics[width=1.5cm,valign=c]{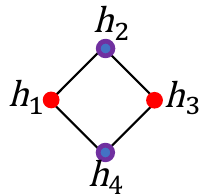})=\frac{\delta_{h_2h_4}}{S_{h_2}}\frac{e^K-1}{\sqrt{q}}+\frac{\delta_{h_1h_3}}{S_{h_1}}
 \notag\\
\end{eqnarray}
are the Boltzmann weights defined on the plaquettes of the medial lattice. It can be found that Eq. \eqref{RSOS_Boltzmann_weight} is exactly equivalent to the last row of Eq. \eqref{edge_tensor}, and $S=S^{(1)}$ ($S=S^{(3)}$) corresponds to the DFib (DYL) case.  Notice that when we contract the tensor network, we add weights $d_i$ to the degrees of freedom which will be contracted. The weights are equivalent to the term $\prod_{i}S_{h_i}$ in Eq. \eqref{RSOS_partition_function}. So an exactly relation between the norms of the PEPS and the partition functions of the RSOS model can be established:
\begin{equation}
  \mathcal{Z}=(qe^{-K})^{\#\text{site}/4}\mathcal{Z}_\text{RSOS},
\end{equation}
where $\mathcal{Z}=\langle\Psi(K)|\Psi(K)\rangle$ in the DFib case and $\mathcal{Z}={}_L\langle\Psi(K)|\Psi(K)\rangle_R$ in the DYL case, and $\#\text{site}$ is the number of sites of the medial lattice.

It has been proved that the partition functions of RSOS models constructed using $S^{(r)}$ are equivalent to that of $q$-state Potts models with $q=4\cos^2(r\pi/p)$\cite{he_2020_geometrical}, where $p$ is the Coxeter number of the Dynkin diagram. So, in the DFib case, the RSOS model is equivalent to the $(\phi+2)$-state Potts model, and this is consistent with the well-known results\cite{fidkowski_string_2009,fendley_topological_2008}. In the DYL case, the RSOS model is equivalent to the $(\phi^\prime+2)$-state Potts model. When $e^K=1+\sqrt{q}$, the RSOS models are critical. The central charges of the CFTs describing the critical points are $c=1-6r^2/[p(p-r)]$, where $p=10$ and $r=1$ ($r=3$) for the DFib (DYL) case.

Moreover, the order parameters of the RSOS models can be defined as $S^{(5)}_{h}/S_h$ and $S^{(\bar{5})}_{h}/S_h$\cite{pasquier_1987_operator,pasquier_Dynkin_RSOS,he_2020_geometrical}, where the denominator $S_h$ cancels the weights assigned previously and the numerator assigns new weights. At the critical points, the scaling dimensions of the RSOS order parameters are given by
\begin{equation}
  \Delta_{\sigma}=\frac{(p/2)^2-r^2}{2p(p-r)}.
\end{equation}
So in the DFib (DYL) case $\Delta_\sigma=2/15$ ($4/35$). In addition, although it is tricky to talk about the definition of Potts order parameters when $q$ is not an integer, the scaling dimensions $\Delta_\sigma$ and $\Delta_\epsilon$ of the Potts models are known as a function of $q$\cite{Potts_scaling_dims}:
\begin{equation}\label{Potts_scaling_dims_formula}
  \Delta_\sigma=\frac{\delta^2-4}{8\delta(\delta-1)},\quad \Delta_\epsilon=\frac{\delta+2}{2(\delta-1)},
\end{equation}
where $q=4\cos^2(\pi/\delta)$. So we say that the scaling dimensions of the RSOS order parameters coincide with those of the Potts order parameters.

Furthermore,
because in the procedures of mapping tensor networks to the RSOS models, the virtual loops around the plaquettes remain unchanged and they are equivalent to the heights on the dual lattice (one sublattice of the medial lattice). It can therefore be proved that the expectation values of the RSOS order parameters are equivalent to the condensate fractions. At first according to Eq. \eqref{end_tensor}, one can check that
\begin{equation}\label{E_tau_1}
  E_{\bm{b}1}=d_\tau^{1/2}\delta_{uv}\delta_{xy}d_v^{-1/4}d_u^{-1/4}d_w^{-2}(-1)^{\delta_{w\tau}}.
\end{equation}
The term $d_v^{-1/4}d_u^{-1/4}$ cancels the weights assigned to $u$ and $v$, which arise from the convention of contraction. The term $d_w^{-2}(-1)^{\delta_{w\tau}}$ cancels the weight $d_w$ previously assigned to virtual loop $w$ and assigns a new weight $(-1)^{\delta_{w\tau}}/d_w$ to $w$. So inserting $E_{\bm{b}1}$ into the PEPS just locally modifies the weight on virtual loops $w$ from $(1,d_\tau)$ to $(1,-1/d_\tau)$. Taking the loop $w^\prime$ in another layer into consideration, for the DFib case we replace the original weights $(1,d_\tau)\otimes (1,d_\tau)=(1,d_\tau,d_\tau,d_\tau^2)$ with
\begin{eqnarray}
  (1,d_\tau)\otimes(1,-\frac{1}{d_\tau})&=&(S^{(5)}_1,S^{(5)}_5,S^{(5)}_6,S^{(5)}_3)\quad \text{or}\notag\\
  (1,-\frac{1}{d_\tau})\otimes(1,d_\tau)&=&(S^{(\bar{5})}_1,S^{(\bar{5})}_5,S^{(\bar{5})}_6,S^{(\bar{5})}_3),\notag
\end{eqnarray}
and for the DYL case we replace the original weights with
\begin{eqnarray}
(1,d_\tau)\otimes(1,-\frac{1}{d_\tau})&=&(S^{(\bar{5})}_1,S^{(\bar{5})}_5,S^{(\bar{5})}_6,S^{(\bar{5})}_3)\quad \text{or}\notag\\
(1,-\frac{1}{d_\tau})\otimes(1,d_\tau)&=&(S^{(5)}_1,S^{(5)}_5,S^{(5)}_6,S^{(5)}_3).\notag
\end{eqnarray}
Because of the above equations, we can therefore identify that inserting the $E_{\bm{b}1}$ shown in \eqref{E_tau_1} into the tensor networks is exactly equivalent to evaluating order parameters $S^{(5)}_{h}/S_h$ and $S^{(\bar{5})}_{h}/S_h$. Therefore the condensate fractions are exactly equivalent to the expectation values of RSOS order parameters:
\begin{equation}
\mathcal{F}_{\bm{1}}^{\bm{b}}=\mathcal{F}_{\bm{b}}^{\bm{1}}\propto\left\langle \frac{S^{(5)}_{h_i}}{S_{h_i}}\right\rangle_{\mathcal{Z}_\text{RSOS}}=
\left\langle \frac{S^{(5^\prime)}_{h_i}}{S_{h_i}}\right\rangle_{\mathcal{Z}_\text{RSOS}},
\end{equation}
where $\langle\bullet\rangle_{\mathcal{Z}_{\text{RSOS}}}$ denotes the thermal ensemble average.
\section{Topological sectors of the transfer operator and form factors of the correlation functions}\label{corr_form_factor}

 The form factors together with the transfer operators determine the correlation lengths
 of different kinds of correlation functions. On an infinitely long cylinder, the repeating units of the norms of the PEPS in Figs. \ref{Fig_Tensor} (a) and (b) are the transfer operators $\mathbb{T}_{1}^{1}$ and $\mathbb{T}_{\tau}^{\tau}$ with the closed boundary conditions, see Fig. \ref{figure_double_layer} (b).
 Since the PEPS in Fig.\ref{Fig_Tensor} (a) and (b) are equal to $|\bm{1}\rangle+|\bm{b}\rangle$ and
 $|\bm{b}\rangle+|\bm{\tau}\rangle+|\bm{\bar{\tau}}\rangle$ individually, their transfer operators
 $\mathbb{T}_{1}^{1}$ and $\mathbb{T}_{\tau}^{\tau}$ contain 4 and 9 topological sectors seperately. These topological sectors
 are subblock transfer operators of the MES overlaps: $\langle\bm{\alpha}|\bm{\beta}\rangle$, where
 $\bm{\alpha},\bm{\beta}=\bm{1},\bm{\tau},\bar{\bm{\tau}},\bm{b}$. Using the idempotents in Eq. \eqref{Idempotents}, the projectors
 $ \mathbb{P}_{\bm{\alpha}}^{\bm{\beta}}=P_{\bm{\beta}1}\otimes P_{\bm{\alpha}1}$ and $\mathbb{\tilde{P}}_{\bm{\alpha}}^{\bm{\beta}}=P_{\bm{\bar{\beta}}\tau}\otimes P_{\bm{\alpha}\tau}$ (notice that the idempotents acting on bra and ket layers should have opposite chiralities) can be defined, the superscript (subscript)of $\mathbb{P}_{\bm{\alpha}}^{\bm{\beta}}$ stands for the MES in the bra (ket) layer. And by using the $\mathbb{P}_{\bm{\alpha}}^{\bm{\beta}}$, the subblocks of the trasfer operators can be projected out:
 \begin{eqnarray}\label{block_diagonal}
 \mathbb{P}_{\bm{\alpha}}^{\bm{\beta}}\mathbb{T}_{1}^{1} \mathbb{P}_{\bm{\alpha}}^{\bm{\beta}}&=&\mathbb{T}_{1}^{1} \mathbb{P}_{\bm{\alpha}}^{\bm{\beta}}=\mathbb{P}_{\bm{\alpha}}^{\bm{\beta}}\mathbb{T}_{1}^{1}
,\quad \bm{\alpha},\bm{\beta}=\bm{1},\bm{b}\notag\\
  \mathbb{\tilde{P}}_{\bm{\alpha}}^{\bm{\beta}}\mathbb{T}_{\tau}^{\tau}\mathbb{\tilde{P}}_{\bm{\alpha}}^{\bm{\beta}}&=&\mathbb{T}_{\tau}^{\tau}\mathbb{\tilde{P}}_{\bm{\alpha}\tau}^{\bm{\beta}\tau}
  =\mathbb{\tilde{P}}_{\bm{\alpha}}^{\bm{\beta}}\mathbb{T}_{\tau}^{\tau}, \quad \bm{\alpha},\bm{\beta}=\bm{b},\bm{\tau},\bm{\bar{\tau}}.
\end{eqnarray}

So the eigenvalues and eigenvectors of $\mathbb{T}_1^1$ and $\mathbb{T}_\tau^\tau$ can be classified into different topological sectors. By diagonalizing the transfer operators, we have
 \begin{eqnarray}
   \mathbb{T}^{1}_{1}&=&\sum_{\bm{\alpha},\bm{\beta}=\bm{1},\bm{b}}\sum_{i}\lambda_{\bm{\alpha},i}^{\bm{\beta}}|r_{\bm{\alpha},i}^{\bm{\beta}})(l_{\bm{\alpha},i}^{\bm{\beta}}|,\\
   \mathbb{T}^{\tau}_{\tau}&=&\sum_{\bm{\alpha},\bm{\beta}=\bm{b},\bm{\tau},\bm{\bar{\tau}}}\sum_{i}\lambda_{\bm{\alpha},i}^{\bm{\beta}}|\tilde{r}_{\bm{\alpha},i}^{\bm{\beta}})(\tilde{l}_{\bm{\alpha},i}^{\bm{\beta}}|,
 \end{eqnarray}
where $\lambda_{\bm{\alpha},j}^{\bm{\beta}}$ is the $(j+1)$-th dominant eigenvalue
of the topological sector
$\langle\bm{\alpha}|\bm{\beta}\rangle$ (the spectrum is rescaled such that $\lambda_{\bm{1},0}^{\bm{1}}=1$ for convenience), and the left and right eigenvectors are bi-orthogonal:
\begin{equation}
  (l_{\bm{\alpha},i}^{\bm{\beta}}|r_{\bm{\alpha}^{\prime},j}^{\bm{\beta}^{\prime}})
 =\delta_{\bm{\alpha}\bm{\alpha}^{\prime}}\delta_{\bm{\beta}\bm{\beta}^{\prime}}\delta_{ij},\quad(\tilde{l}_{\bm{\alpha},i}^{\bm{\beta}}|\tilde{r}_{\bm{\alpha}^{\prime},j}^{\bm{\beta}^{\prime}})
 =\delta_{\bm{\alpha}\bm{\alpha}^{\prime}}\delta_{\bm{\beta}\bm{\beta}^{\prime}}\delta_{ij}\notag.
\end{equation}
The dominant eigenvalue $\lambda_{\bm{\alpha},0}^{\bm{\beta}}$ of each topological sector is non-degenerate. And the eigenvalues $\lambda_{\bm{b},i}^{\bm{b}}$ from
  $\mathbb{T}_1^1$ and $\lambda_{\bm{b},i}^{\bm{b}}$ from $\mathbb{T}_\tau^\tau$ are equal, as shown in Figs. \ref{DFib_CFT_spec} (a) and (b) (also in Figs. \ref{DYL_CFT_spec} (a) and (b)).

On an infinitely long cylinder, the correlation function \eqref{cond_corr} can be expressed as:
\begin{eqnarray}
\mathcal{C}_{\bm{b}}^{\bm{1}}(m)
&=&( l_{\bm{1},0}^{\bm{1}}|(\mathbb{E}^{\bm{1}}_{\bm{b}})^\text{T}(\mathbb{T}^{1}_{1})^{m}\mathbb{E}^{\bm{1}}_{\bm{b}}|r_{\bm{1},0}^{\bm{1}})\notag\\
&=&\sum_{\bm{\alpha},\bm{\beta}=\bm{1},\bm{b}}\sum_{i}(\lambda_{\bm{\alpha},i}^{\bm{\beta}})^m(l_{\bm{1},0}^{\bm{1}}|(\mathbb{E}^{\bm{1}}_{\bm{b}})^\text{T}|r_{\bm{\alpha},i}^{\bm{\beta}})(l_{\bm{\alpha},i}^{\bm{\beta}}|\mathbb{E}^{\bm{1}}_{\bm{b}}|r_{\bm{1},0}^{\bm{1}}),\notag
\end{eqnarray}
where $\mathbb{E}^{\bm{1}}_{\bm{b}}$ is obtained by inserting an end tensor $E_{\bm{b}1}$ into the ket layer of the transfer operator $\mathbb{T}^{1}_{1}$, see Fig. \ref{figure_double_layer} (c).
 The denominator of Eq. \eqref{cond_corr} disappears due to $\lambda_{\bm{1},0}^{\bm{1}}=1$.
 Because the $\mathbb{E}^{\bm{1}}_{\bm{b}}|r_{\bm{1},0}^{\bm{1}})$ is a vector belonging to the topological sector $\langle\bm{1}|\bm{b}\rangle$,
  the form factors of the correlation function satisfies:
\begin{equation}\label{form factor}
  (l_{\bm{\alpha},i}^{\bm{\beta}}|\mathbb{E}^{\bm{1}}_{\bm{b}}|r_{\bm{1},0}^{\bm{1}})
  \propto\delta_{\bm{\beta1}}\delta_{\bm{\alpha b}}.
\end{equation}

Therefore, the correlation function can be further simplified:
\begin{eqnarray}
  \mathcal{C}_{\bm{b}}^{\bm{1}}(m)&=&\sum_{i}(\lambda_{\bm{b},i}^{\bm{1}})^m(l_{\bm{1},0}^{\bm{1}}|(\mathbb{E}^{\bm{1}}_{\bm{b}})^\text{T}|r_{\bm{b},i}^{\bm{1}})(l_{\bm{b},i}^{\bm{1}}|\mathbb{E}^{\bm{1}}_{\bm{b}}|r_{\bm{1},0}^{\bm{1}})\notag\\
  &\approx&(\lambda_{\bm{b},0}^{\bm{1}})^m(l_{\bm{1},0}^{\bm{1}}|(\mathbb{E}^{\bm{1}}_{\bm{b}})^\text{T}|r_{\bm{b},0}^{\bm{1}})(l_{\bm{b},0}^{\bm{1}}|\mathbb{E}^{\bm{1}}_{\bm{b}}|r_{\bm{1},0}^{\bm{1}})\notag\\
  &\propto& \exp(-m/\xi_{\bm{b}}^{\bm{1}}),
\end{eqnarray}
where the approximation in the second line is valid for large $m$ and $1/\xi_{\bm{b}}^{\bm{1}}=-\log(\lambda_{\bm{b},0}^{\bm{1}})$. At the critical point, the eigenvalues of the transfer operators are also related to the scaling dimensions of CFT \cite{cardy_1986},
\begin{equation}\label{cardy_relation}
  \Delta_{\bm{\alpha},j}^{\bm{\beta}}=-L_y\log\left(\lambda_{\bm{\alpha},j}^{\bm{\beta}}\right)/(2\pi)+\mathcal{O}(L_y^{-\gamma_{\bm{\alpha},j}^{\bm{\beta}}}),
\end{equation}
where $L_y$ is the circumference of the transfer operators, $\gamma_{\bm{\alpha},j}^{\bm{\beta}}>0$ and the finite size corrections vanish with $L_y$ increasing, as shown in Figs. \ref{DFib_CFT_spec} (c) and (d) (also in Figs. \ref{DYL_CFT_spec} (c) and (d)). Hence we can numerically identify the scaling dimension $\Delta_\sigma=\Delta_{\bm{b},0}^{\bm{1}}$.

The correlation function \eqref{conf_corr} can be written as
\begin{eqnarray}
\mathcal{C}_{\bm{\tau}}^{\bm{\tau}}(m)&=&
(l_{\bm{1},0}^{\bm{1}}|(\mathbb{E}^{\bm{\bar{\tau}}}_{\bm{\bar{\tau}}})^\text{T}(\mathbb{T}^{\tau}_{\tau})^{m}\mathbb{E}^{\bm{\tau}}_{\bm{\tau}}|r_{\bm{1},0}^{\bm{1}})\notag\\
&=&\sum_{\substack{\bm{\alpha},\bm{\beta}=\\\bm{b},\bm{\tau},\bm{\bar{\tau}}}}
  \sum_{i}(\lambda_{\bm{\alpha},i}^{\bm{\beta}})^m(l_{\bm{1},0}^{\bm{1}}|
  (\mathbb{E}^{\bm{\bar{\tau}}}_{\bm{\bar{\tau}}})^\text{T}|\tilde{r}_{\bm{\alpha},i}^{\bm{\beta}})
  (\tilde{l}_{\bm{\alpha},i}^{\bm{\beta}}|\mathbb{E}^{\bm{\tau}}_{\bm{\tau}}|r_{\bm{1},0}^{\bm{1}})\notag,
\end{eqnarray}
where $\mathbb{E}^{\bm{\tau}}_{\bm{\tau}}$ is obtained by inserting the end tensors $E_{\bm{\bar{\tau}}\tau}$ and $E_{\bm{\tau}\tau}$ into the bra and ket layers of transfer operator $\mathbb{T}_1^1$, see Fig. \ref{figure_double_layer} (d).  Because the form factors have the following property:
\begin{equation}
  (\tilde{l}_{\bm{\alpha},i}^{\bm{\beta}}|\mathbb{E}^{\bm{\tau}}_{\bm{\tau}}|r_{\bm{1},0}^{\bm{1}})
  \propto\delta_{\bm{\alpha\tau}}\delta_{\bm{\beta\tau}}.
\end{equation}
The corrlation function can be simplified for the sufficient large $m$:
\begin{eqnarray}
  \mathcal{C}_{\bm{\tau}}^{\bm{\tau}}(m)&=&
  \sum_{i}(\lambda_{\bm{\tau},i}^{\bm{\tau}})^m(l_{\bm{1},0}^{\bm{1}}|
  (\mathbb{E}^{\bm{\bar{\tau}}}_{\bm{\bar{\tau}}})^\text{T}|\tilde{r}_{\bm{\tau},i}^{\bm{\tau}})
  (\tilde{l}_{\bm{\tau},i}^{\bm{\tau}}|\mathbb{E}^{\bm{\tau}}_{\bm{\tau}}|r_{\bm{1},0}^{\bm{1}})\notag\\
  &\approx&
 (\lambda_{\bm{\tau},0}^{\bm{\tau}})^m(l_{\bm{1},0}^{\bm{1}}|
  (\mathbb{E}^{\bm{\bar{\tau}}}_{\bm{\bar{\tau}}})^\text{T}|\tilde{r}_{\bm{\tau},0}^{\bm{\tau}})
  (\tilde{l}_{\bm{\tau},0}^{\bm{\tau}}|\mathbb{E}^{\bm{\tau}}_{\bm{\tau}}|r_{\bm{1},0}^{\bm{1}})\notag\\
  &\propto& \exp(-m/\xi_{\bm{\tau}}^{\bm{\tau}}),
\end{eqnarray}
 where $1/\xi_{\bm{\tau}}^{\bm{\tau}}=-\log(\lambda_{\bm{\tau},0}^{\bm{\tau}})$
 at the critical point. According to Eq. \eqref{cardy_relation}, we have $\Delta_\mu=\Delta_{\bm{\tau},0}^{\bm{\tau}}$.

The two terms in the correlation function \eqref{trivial_corr} can be written as:
\begin{eqnarray}
 \frac{\langle\bm{1}|\sigma_0^z\sigma_m^z|\bm{1}\rangle}{\langle\bm{1}|\bm{1}\rangle}
 &=&\sum_{i}(\lambda_{\bm{1},i}^{\bm{1}})^{m}(l_{\bm{1},0}^{\bm{1}}|(\mathbb{E}^{\bm{1}}_{\bm{1}})^\text{T}|r_{\bm{1},i}^{\bm{1}})(l_{\bm{1},i}^{\bm{1}}|\mathbb{E}^{\bm{1}}_{\bm{1}}|r_{\bm{1},0}^{\bm{1}}),\notag\\
 \frac{\langle\bm{1}|\sigma_j^z|\bm{1}\rangle}{\langle\bm{1}|\bm{1}\rangle}&=&(l_{\bm{1},0}^{\bm{1}}|\mathbb{E}^{\bm{1}}_{\bm{1}}|r_{\bm{1},0}^{\bm{1}}).
\end{eqnarray}
 where $\mathbb{E}^{\bm{1}}_{\bm{1}}$ is obtained by inserting a $\sigma^z$ operator in a physical degree of freedom of the $\mathbb{T}^{1}_{1}$. Because the form factors are non-zero only for the trivial sector:
\begin{equation}
  (l_{\bm{\alpha},i}^{\bm{\beta}}|\mathbb{E}^{\bm{1}}_{\bm{1}}|r_{\bm{1},0}^{\bm{1}})
  \propto\delta_{\bm{\beta1}}\delta_{\bm{\alpha 1}},
\end{equation}
the correlation function \eqref{trivial_corr} can be approximated as
\begin{eqnarray}
  \mathcal{C}_{\bm{1}}^{\bm{1}}(m)&\approx&(\lambda_{\bm{1},1}^{\bm{1}})^{m}(l_{\bm{1},0}^{\bm{1}}
  |(\mathbb{E}^{\bm{1}}_{\bm{1}})^\text{T}|r_{\bm{1},1}^{\bm{1}})(l_{\bm{1},1}^{\bm{1}}|\mathbb{E}^{\bm{1}}_{\bm{1}}|r_{\bm{1},0}^{\bm{1}})\notag\\
  &\propto&\exp(-m/\xi_{\bm{1}}^{\bm{1}})
\end{eqnarray}
for the sufficiently large $m$, and $1/\xi_{\bm{1}}^{\bm{1}}=-\log(\lambda_{\bm{1},1}^{\bm{1}})$. Again from Eq. \eqref{cardy_relation}, we have $\Delta_\epsilon=\Delta_{\bm{1},1}^{\bm{1}}$.
Notice that the correlation length is determined by the second largest eigenvalue of the trivial sector instead of the first one.

\section{Numerical methods}\label{Numerical_method}
In this appendix, the main ideas of the numerical methods are briefly introduced.
The double tensor generating the wavefunction norm is obtained by contracting the physical degrees of freedom of a local tensor and its complex conjugate:
\begin{equation}
 \includegraphics[width=3cm,valign=c]{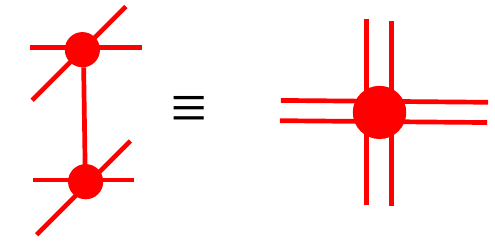}.
\end{equation}
In the VUMPS method, the dominant eigenvector of the transfer operator are approximated by an iMPS:
\begin{equation}
 \includegraphics[width=4cm,valign=c]{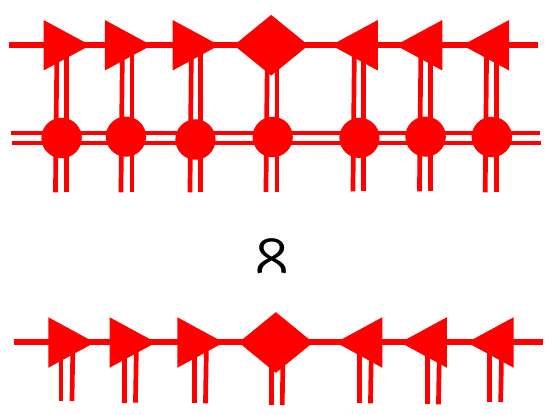},
\end{equation}
where the triangles on the left and right sides as well as the diamond in the center are tensors of mixed canonical gauge iMPS, which can be optomized using the VUMPS algorithm\cite{VUMPS_2019,Verstraete_corner_2018}. Define the left and right fixed points of the channel operators:
\begin{equation}\label{channel_operator}
 \includegraphics[width=7cm,valign=c]{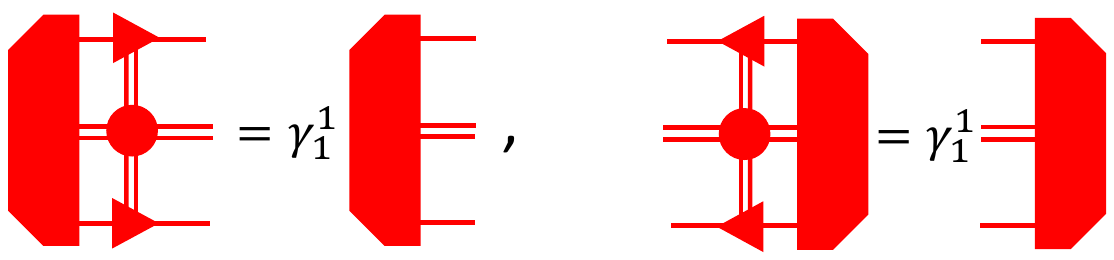},
\end{equation}
where the MPS tensors in the bottom are complex conjugates of those in the top.
The condensate fraction can be expressed using these fixed points:
\begin{equation}
 \mathcal{F}_{\bm{b}}^{\bm{1}}=\includegraphics[width=4cm,valign=c]{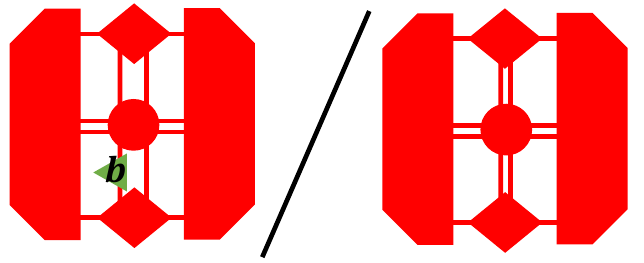}.
\end{equation}
And the correlation function \eqref{cond_corr} is
\begin{equation}
 \mathcal{C}_{\bm{b}}^{\bm{1}}(|i-j|)=\includegraphics[width=6cm,valign=c]{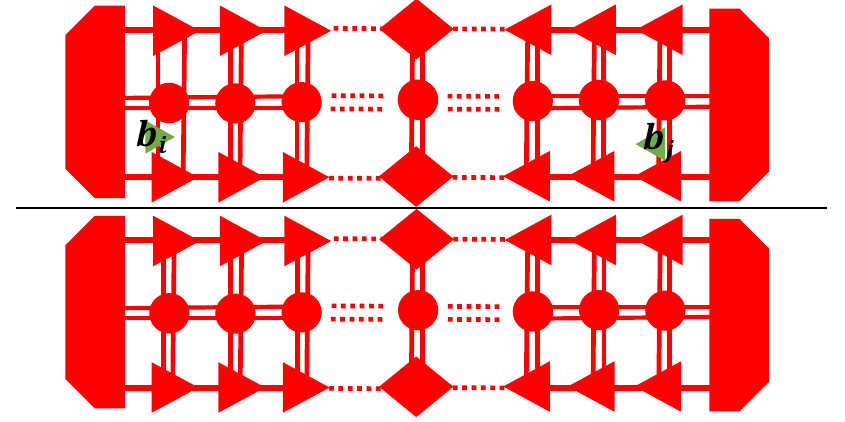}.
\end{equation}
To compute the confinement fraction,
define a channel operator inserted with the tensor product of two MPO tensors ($n=\tau$) \eqref{MPO_tensor}
and find its left fixed point:
\begin{equation}\label{channel_operator_with_MPO}
\includegraphics[width=3.5cm,valign=c]{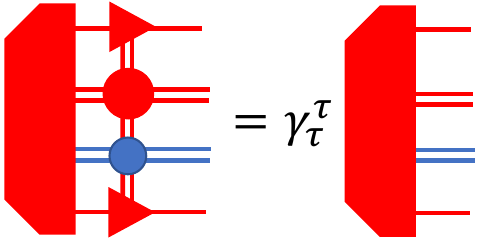}.
\end{equation}
Then the deconfinement fraction can be represented as
\begin{equation}
\mathcal{F}_{\bm{\tau}}^{\bm{\tau}}=\includegraphics[width=4.5cm,valign=c]{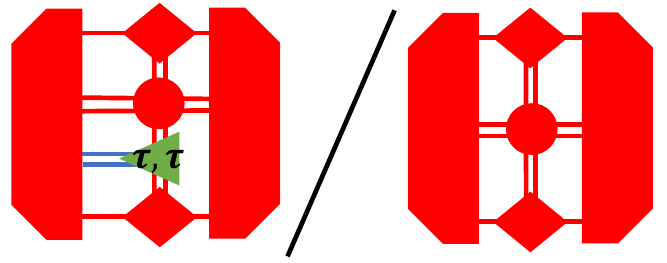}\times\lim_{L_x\rightarrow\infty}
  (\frac{\gamma_{\tau}^{\tau}}{\gamma_{1}^{1}})^{L_x},
\end{equation}
where the green double layer end point tensor is a tensor product of the $E_{\bm{\tau}\tau}$
and $E_{\bar{\bm{\tau}}\tau}$ defined in \eqref{end_tensor}.
Actually, the confinement length $\xi_{\bm{\tau}}^{\bm{\tau}}=1/\log(|\gamma_1^1/\gamma_{\tau}^{\tau}|)$,
where $\gamma_{1}^{1}$ and $\gamma_{\tau}^{\tau}$ are defined
in Eqs. \eqref{channel_operator} and  \eqref{channel_operator_with_MPO}, separately.
In the topological phase, the iMPS respects the MPO symmetry,
so that $\gamma_{1}^{1}=\gamma_{\tau}^{\tau}=1$ and $\xi_{\bm{\tau}}^{\bm{\tau}}=+\infty$.
In the non-topological phase, the iMPS spontaneously breaks the MPO symmetry and
$\gamma_{1}^{1}>\gamma_{\tau}^{\tau}$, resulting in a finite confinement length.

These physical quantities can also be calculated using the CTM method,
especially in the DYL case where the transfer operator is non-Hermitian.
The environments of an infinitely large tensor network of wavefunction norm can be approximated by CTMs:
\begin{equation}
\includegraphics[width=6cm,valign=c]{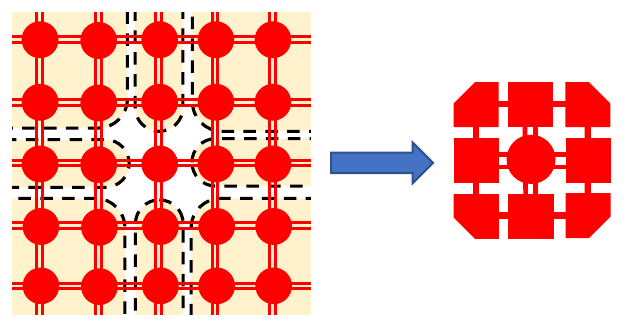}.
\end{equation}
There are four corner tensors and four edge tensors obtained from
the CTMRG optimization\cite{corboz_corner_2014}.
In the CTM framework, the condensate fraction is
\begin{equation}
  \mathcal{F}_{\bm{b}}^{\bm{1}}=\includegraphics[width=4cm,valign=c]{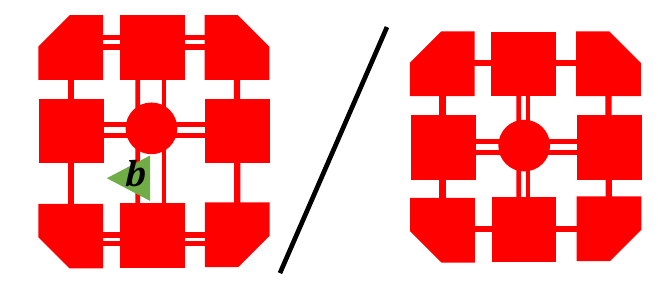}.
\end{equation}
Because the left and right parts of the environments are the left and right fixed points
of the channel operator:
\begin{equation}\label{CTM_left_fixed_point}
\includegraphics[width=7cm,valign=c]{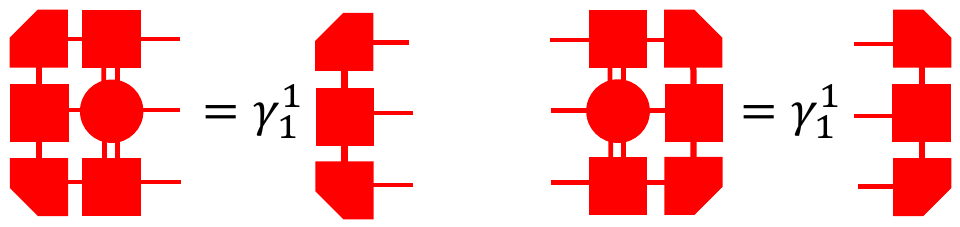},
\end{equation}
The correlation function can be estimated by:
\begin{equation}
 \mathcal{C}_{\bm{b}}^{\bm{1}}(|i-j|)=\includegraphics[width=6cm,valign=c]{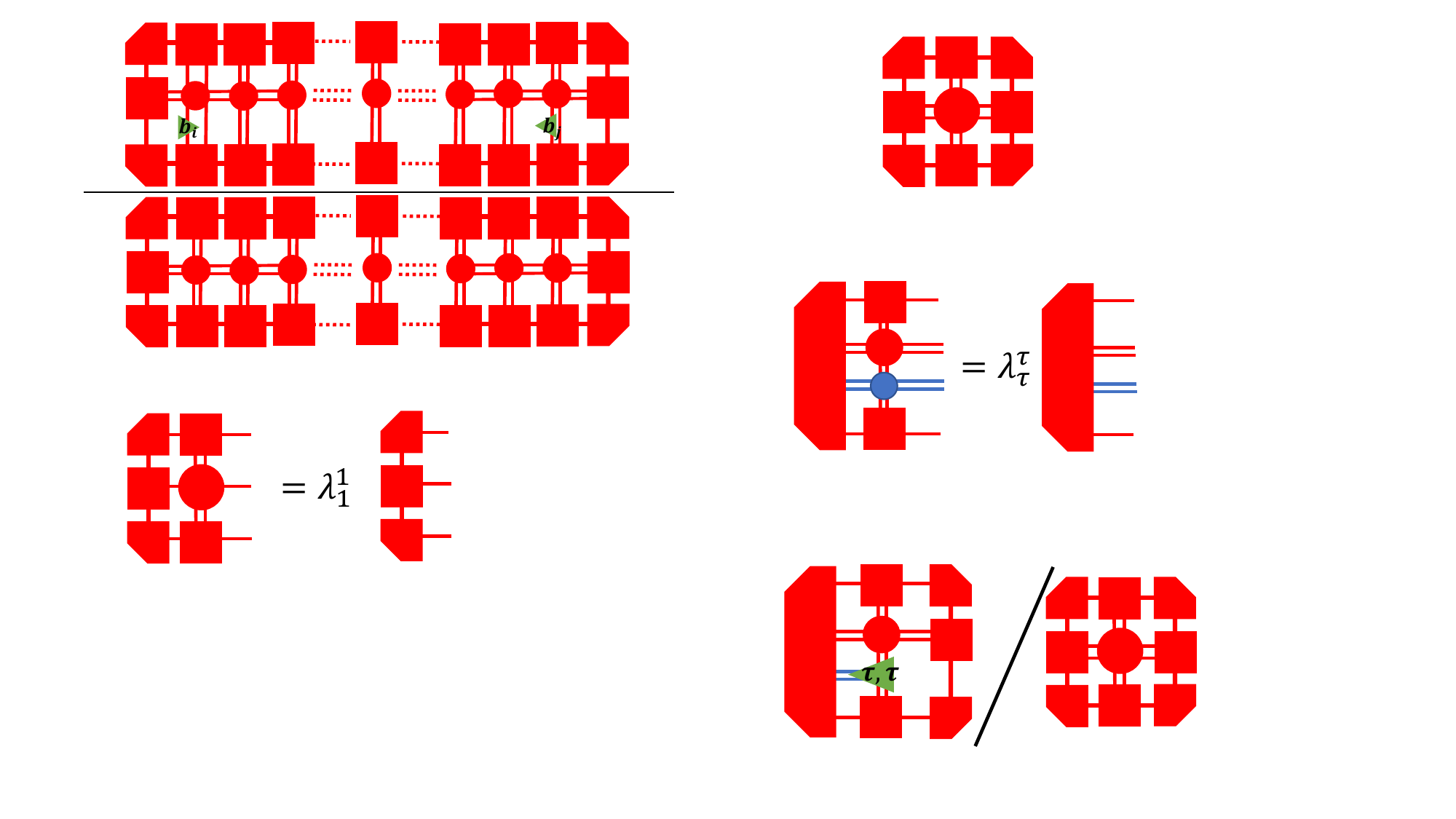}.
\end{equation}
Define a channel
operator inserted with the tensor product of two non-trivial MPO
tensors in Eq. (32) and find its left fixed point:
\begin{equation}\label{CTM_left_fixed_point_with_MPO}
 \includegraphics[width=3.5cm,valign=c]{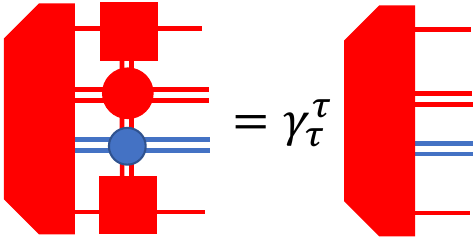},
\end{equation}
then the deconfinement fraction can be expressed as:
\begin{equation}
\mathcal{F}_{\bm{\tau}}^{\bm{\tau}}=\includegraphics[width=4cm,valign=c]{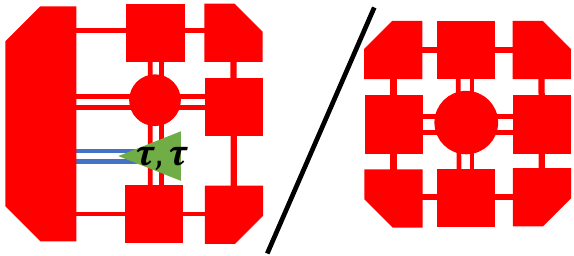}\times\lim_{L_x\rightarrow\infty}
  (\frac{\gamma_{\tau}^{\tau}}{\gamma_{1}^{1}})^{L_x}.
\end{equation}
Then confinement length $\xi_{\bm{\tau}}^{\bm{\tau}}=1/\log(|\gamma_1^1/\gamma_{\tau}^{\tau}|)$, where $\gamma_{1}^{1}$ and $\gamma_{\tau}^{\tau}$ are defined in Eqs. \eqref{CTM_left_fixed_point} and  \eqref{CTM_left_fixed_point_with_MPO}, separately.
\bibliography{refs}

\end{document}